\documentclass[conference]{IEEEtran}
\IEEEoverridecommandlockouts

\usepackage{cite}
\usepackage{amsmath,amssymb,amsfonts}
\usepackage{graphicx}
\usepackage{subcaption}
\usepackage{textcomp}
\usepackage[table]{xcolor}
\usepackage{url}
\usepackage{hyperref}
\usepackage{tikz}
\usepackage{booktabs}

\usepackage{multirow}

\usepackage[ruled,vlined,linesnumbered]{algorithm2e}

\usepackage{xspace}
\usepackage[capitalize,sort&compress]{cleveref}

\usepackage{listings}

\definecolor{pinkshocking}{rgb}{0.99,0.01,0.84}

\definecolor{mGreen}{rgb}{0,0.6,0}
\definecolor{mGray}{rgb}{0.5,0.5,0.5}
\definecolor{mPurple}{rgb}{0.58,0,0.82}
\definecolor{backgroundColour}{rgb}{0.99,0.99,0.99}

\definecolor{mGrayBox}{HTML}{e0e0e0}

\newcommand{\ie}{\textit{i.e.,} }
\newcommand{\eg}{\textit{e.g.,} }

\newcommand{\sys}{\textsf{FishFuzz}\xspace}

\usepackage[inline]{enumitem}

\newcommand{\circled[1]}{\tikz[baseline=(char.base)]{\node[font=\sffamily,
		shape=circle,draw,inner sep=0.5pt,color=black,fill=white] (char) {#1};}}

\definecolor{interexplorationcolor}{HTML}{daf0f8}
\definecolor{intraexplorationcolor}{HTML}{b2d8b2}
\definecolor{exploitationcolor}{HTML}{ffb2b2}

\newcommand{\exploitationsquare}{\fcolorbox{black}{exploitationcolor}{\rule{0pt}{4pt}\rule{4pt}{0pt}}}
\newcommand{\interexplorationsquare}{\fcolorbox{black}{interexplorationcolor}{\rule{0pt}{4pt}\rule{4pt}{0pt}}}
\newcommand{\intraexplorationsquare}{\fcolorbox{black}{intraexplorationcolor}{\rule{0pt}{4pt}\rule{4pt}{0pt}}}

\graphicspath{{./import/fig/}}

\newcommand{\worstasancovaflp}{$-4.9\%$\xspace}
\newcommand{\bestasancovaflp}{$47.95\%$\xspace}
\newcommand{\bestasancovtort}{$58.53\%$\xspace}
\newcommand{\bestasancovparm}{$132.02\%$\xspace}

\newcommand{\worstasanreaaflp}{$-3.88\%$\xspace}
\newcommand{\bestasanreaaflp}{$40.23\%$\xspace}
\newcommand{\bestasanreatort}{$38.15\%$\xspace}
\newcommand{\bestasanreaparm}{$116.63\%$\xspace}

\newcommand{\bestubsancovsav}{$21.33\%$\xspace}

\newcommand{\bestubsanreasav}{$19.61\%$\xspace}

\newcommand{\totbugasanbest}{$40$\xspace}
\newcommand{\totbugasanavg}{$30.6$\xspace}

\newcommand{\totbugasanbestaflp}{$20$\xspace}
\newcommand{\totbugasanavgaflp}{$16$\xspace}

\newcommand{\totbugasanbesttort}{$20$\xspace}
\newcommand{\totbugasanavgtort}{$16.2$\xspace}

\newcommand{\totbugasanbestparm}{$9$\xspace}
\newcommand{\totbugasanavgparm}{$9$\xspace}

\newcommand{\maxbestubsantrg}{$62.50\%$\xspace}
\newcommand{\minbestubsantrg}{$8.21\%$\xspace}

\newcommand{\maxavgubsantrg}{$52.17\%$\xspace}
\newcommand{\minavgubsantrg}{$2.99\%$\xspace}

\newcommand{\totalbugsrepr}{$47$\xspace}
\newcommand{\totalbugsreprwin}{$33$\xspace}
\newcommand{\totalbugsreprdraw}{$6$\xspace}
\newcommand{\totalbugsreprloose}{$8$\xspace}

\newcommand{\totalbugsreprwinper}{$70.2\%$\xspace}
\newcommand{\totalbugsreprdrawper}{$12.7\%$\xspace}
\newcommand{\totalbugsreprlooseper}{$17.02\%$\xspace}

\newcommand{\numnewcvesfound}{$18$\xspace}
\newcommand{\numnewbugsfound}{$25$\xspace}

\newcommand{\numrealworldprograms}{$28$\xspace}
\newcommand{\numasanprograms}{$9$\xspace}
\newcommand{\numubsanprograms}{$7$\xspace}
\newcommand{\numtotalprograms}{$44$\xspace}

\newcommand{\numbuglessthanthreeday}{$12$\xspace}
\newcommand{\numbugmorethanthreeday}{$2$\xspace}
\newcommand{\numberofasanbugs}{$11$\xspace}
\newcommand{\numberofubsanbugs}{$3$\xspace}

\newcommand{\numdividezerobugs}{$1$\xspace}
\newcommand{\numdheapoverbugs}{$8$\xspace}
\newcommand{\numassertbugs}{$3$\xspace}
\newcommand{\numstackexbugs}{$1$\xspace}
\newcommand{\numshiftexpbugs}{$1$\xspace}

\newcommand{\maxtargetscovered}{$20k$\xspace}
\newcommand{\numbenchmark}{$three$\xspace}

\newcommand{\maxcoveragedelta}{$132\%$\xspace}
\newcommand{\maxtargetsdelta}{$116\%$\xspace}
\newcommand{\maxbugsdelta}{$2$x\xspace}
\newcommand{\maxtargettriggerdelta}{$62\%$\xspace}

\newcommand{\numtotalbugsff}{$45$\xspace}
\newcommand{\numbugsonlyaflpp}{$1$\xspace}
\newcommand{\numbugsonlyparm}{$1$\xspace}

\newcommand{\numtotaltargetsff}{$405$\xspace}
\newcommand{\numtargetsonlysavior}{$27$\xspace}

\newcommand{\qsymtrigger}{$64.98\%$\xspace}
\newcommand{\qsymreach}{$116.07\%$\xspace}
\newcommand{\qsymcov}{$41.26\%$\xspace}
\newcommand{\qsympath}{$91.54\%$\xspace}

\def\BibTeX{{\rm B\kern-.05em{\sc i\kern-.025em b}\kern-.08em
    T\kern-.1667em\lower.7ex\hbox{E}\kern-.125emX}}
\begin{document}

\date{}

\title{\sys: Throwing Larger Nets to Catch Deeper Bugs}


\author{
    \IEEEauthorblockN{
      Han Zheng\IEEEauthorrefmark{1}, Jiayuan Zhang\IEEEauthorrefmark{2}\IEEEauthorrefmark{1}, Yuhang Huang\IEEEauthorrefmark{1}, Zezhong Ren\IEEEauthorrefmark{1}, He Wang\IEEEauthorrefmark{3}
    }
		\IEEEauthorblockN{
			Chunjie Cao\IEEEauthorrefmark{4}, Yuqing Zhang\IEEEauthorrefmark{1}\IEEEauthorrefmark{3}\IEEEauthorrefmark{4}, Flavio Toffalini\IEEEauthorrefmark{5}, Mathias Payer\IEEEauthorrefmark{5}
		}

    \IEEEauthorblockA{\IEEEauthorrefmark{1} National Computer Network Intrusion Protection Center, University of Chinese Academy of Sciences, China}
    \IEEEauthorblockA{\IEEEauthorrefmark{2} School of Computer and Communication, Lanzhou University of Technology, China}
    \IEEEauthorblockA{\IEEEauthorrefmark{3} School of Cyber Engineering, Xidian University, China}
    \IEEEauthorblockA{\IEEEauthorrefmark{4} School of Cyberspace Security, Hainan University, China}
    \IEEEauthorblockA{\IEEEauthorrefmark{5} EPFL, Switzerland}
}

\maketitle


\begin{abstract}
Greybox fuzzing is the de-facto standard to discover bugs during
development. Fuzzers execute many inputs to maximize the amount of
reached code.
Recently, Directed Greybox Fuzzers (DGFs) propose an alternative strategy that
goes beyond ``just'' coverage: driving testing toward specific code
targets by selecting ``closer'' seeds.
DGFs go through different phases: exploration (\ie reaching interesting
locations) and exploitation (\ie triggering bugs).
In practice, DGFs leverage coverage to directly measure exploration, while
exploitation is, at best, measured indirectly by alternating between different targets.
Specifically, we observe two limitations in existing DGFs: (i) they lack
precision in their distance metric, \ie averaging multiple paths and targets
into a single score (to decide which seeds to prioritize), and (ii)
they assign energy to seeds in a round-robin fashion without adjusting the
priority of the targets (exhaustively explored targets should be dropped).

We propose \sys, which draws inspiration from trawl fishing: first casting a
wide net, scraping for high coverage, then slowly pulling it in to maximize the harvest.
The core of our fuzzer is a novel seed selection strategy that builds on two
concepts: (i) a novel multi-distance metric whose precision is independent of
the number of targets, and (ii) a dynamic target ranking to automatically
discard exhausted targets.
This strategy allows \sys to seamlessly scale to tens of thousands of
targets and dynamically alternate between exploration and exploitation
phases. We evaluate \sys by leveraging \emph{all} sanitizer labels as targets.
Extensively comparing \sys against modern DGFs and coverage-guided fuzzers shows
that \sys reached higher coverage compared to the direct competitors, reproduces
existing bugs (\totalbugsreprwinper faster), and finally discovers
\numnewbugsfound new bugs (\numnewcvesfound CVEs) in \numtotalprograms programs.

\end{abstract}



\section{Introduction}

Greybox fuzzing is the established technique to automatically test and find bugs
in programs.  The base concept of a fuzzer is as simple as effective: execute
the target program with inputs (seeds), observe its behavior, and report
observed crashes. The seed that triggered the crash allows reproducing the crash
later during debugging. The effectiveness of these fuzzers moved researcher and
companies to invest considerably effort in this technology, thus producing more
sophisticated fuzzing designs~\cite{afl, aflpp, osterlund2020parmesan,
wagner2015high, wang2020not, chen2018angora, chen2018hawkeye, chen2020savior,
48314, godefroid2012sage}.

Greybox fuzzing is coverage-guided, which means that fuzzers aim at
maximizing the amount of explored code~\cite{afl,aflpp,48314}.
However, solely increasing coverage does not guarantee finding all bugs:
once code is reached (covered), the fuzzer
drives inputs toward unexplored regions, thus missing the opportunity to
trigger potential error cases in already code.
Fuzzing encompasses two aspects: \emph{exploration} and \emph{exploitation}.
A fuzzer needs both to explore the program broadly but also trigger bugs in code
it reaches. During exploration, the main goal is to increase coverage. During
exploitation, the main goal is to trigger bugs by executing a piece of code with
diverse inputs. Greybox fuzzers directly measure exploration by tracking newly
reached code areas but \emph{have no feedback for exploitation}. Greybox fuzzers
assume that exploitation is \emph{implicitly covered} by random mutations that
repeatedly execute the same code \emph{accidentally}.

To make exploitation a first class citizen, researchers introduced Directed 
Greybox Fuzzers
(DGF~\cite{bohme2017directed}) which direct exploration towards a specific code
location (target), improving the likelihood to find bugs at that
location~\cite{bohme2017directed, osterlund2020parmesan, chen2018hawkeye,
chen2020savior}.
DGFs leverage the distance between seeds and
targets to fine-tune inputs more likely to trigger errors.

Recent DGFs try to balance \emph{exploitation} and  \emph{exploration} to
automatically trigger larger sets of targets (on the order of
thousands)~\cite{regression, osterlund2020parmesan, chen2018hawkeye, chen2020savior}.
However, we observe two core limitations in previous works.
First, they model the distance between seeds and targets as a single harmonic
average. This means that, regardless of the size of the target set, they always
collapse the distance seeds-targets into a single scalar.
In short, it does not matter what target is reached as long as any target is
reached.
This approach intrinsically limits the precision of the fuzzer since, with
larger target sets, the distance tries to fit (unrelated) targets
contemporaneously. Moreover, current distance metrics are affected by unresolved
indirect jumps.
Second, current work assigns a time-invariant priority to the targets. They use
different techniques (\eg static analysis, sanitizer labels) to infer
error-prone targets \emph{ahead of time}, \ie before the beginning of the
fuzzing session.
However, we observe that priorities of a target change during the fuzzing
campaign and a fuzzer must adjust its strategy accordingly.
For instance, targets that have been hit frequently are ``\emph{well explored}''
and therefore less likely to be buggy. Less explored targets should therefore be
prioritized to uncover new bugs.
Therefore, a time-invariant priority might misclassify the importance of a code
location and waste fuzzing energy.

Observing the \emph{imprecision of current distance metrics} and
\emph{time-invariant target priority},
we propose \sys: a novel DGF that seamlessly scales to tens of thousands of targets
automatically.
\sys builds on two key contributions: (1) our novel multi-distance metric whose
precision is independent of the number of targets and more robust to indirect
jumps, and (2) a dynamic target ranking to automatically discard exhausted
targets and steer fuzzer energy towards (more) promising locations.
The insight behind our approach is analogous to \emph{trawling} (which inspired
our fuzzer's name). After casting a wide net (capturing many possible targets),
the net is closed gradually. During \emph{exploration} our fuzzer tries to reach
as many targets as possible while during \emph{exploitation}, our fuzzer tracks
how well explored each target is.

We use our insights to design a novel multi-stage seed selection strategy that
promotes seeds according to the current phase the fuzzer is in:
\emph{exploration} or \emph{exploitation}.  Unlike previous works, \sys can
easily handle tens thousands of targets without loss of precision (\eg in our
largest target, we cover over \emph{\maxtargetscovered targets}.
Concurrently, \sys automatically discards unfruitful targets and
priorities rarely tested ones.  The combination of these two strategies allows
the fuzzer to reach more targets (during exploration) and then spread its
energy evenly across the discovered targets (during exploitation). Together,
this results in better bug finding capabilities.

Our \sys prototype extends AFL and we compare it
against three modern DGFs (\ie ParmeSan~\cite{osterlund2020parmesan},
TortoiseFuzz~\cite{wang2020not}, and SAVIOR~\cite{chen2020savior}) as well as
against AFL++~\cite{aflpp} and AFL~\cite{afl}.
We conduct experiments over \numbenchmark benchmarks. The first two are 
composed of \numasanprograms and \numubsanprograms programs taken from
TortoiseFuzz~\cite{wang2020not} and SAVIOR~\cite{chen2020savior}, respectively.
For the third one, we select \numrealworldprograms real programs from other top 
tier fuzzing works.
For what concerns \emph{exploration}, our evaluation shows we easily reach an
higher coverage (up to \maxcoveragedelta more) and hit more targets (up to
\maxtargetsdelta more) compared with the state of the art.
In terms of \emph{exploitation}, we show \sys can trigger
\maxtargettriggerdelta more targets and find up to \maxbugsdelta unique bugs
with respect to its competitors.
More precisely, \sys easily reproduces \numtotalbugsff previous bugs, among
which \totalbugsreprwin in less time (\totalbugsreprwinper) compared to
previous works. Moreover, we discover \numnewbugsfound new bugs from which
\numnewcvesfound are already confirmed CVEs.
Additionally, we measure the contribution of \emph{exploration} and
\emph{exploitation} phases. The results show (1) \sys better
balances the fuzzing energy among the targets and (2) we show the impact of the
\emph{exploration} and \emph{exploitation} strategy in terms of coverage and
targets trigger.

To sum up, our contributions are:
\begin{itemize}
	\item \sys: a DGF that employs a novel multi-stage seed selection strategy
	to maximizes the \emph{explored} and \emph{exploited} targets.
	\item A novel multi-distance metric between seeds and targets that is
	independent of the size of the target set and robust against unsolved
	indirect jumps.
	\item A dynamic target ranking that automatically guides fuzzer energy
	towards promising locations, while discarding thoroughly explored ones.
	\item A detailed evaluation against the state of the art and new
	\numnewbugsfound bugs found (\numnewcvesfound CVEs) in \numtotalprograms
	programs.
\end{itemize}

We will release the full source code with the publication of the paper and a
demonstration prototype is available at \url{https://zenodo.org/record/6405418}.


\section{Background}

\sys heavily modifies seed selection and introduces a new distance metric. To
understand the limitations of existing queue culling and distance metrics, we
first introduce the concepts and then highlight challenges in their current
form.


\subsection{Queue Culling}
\label{ssec:queue-culling-bkg}

Modern fuzzers, such as AFL~\cite{afl} and AFL++~\cite{aflpp}, take as
input a program and a set of inputs (seed) to submit to the program.
Their workflow is loop-based: they select seeds, mutated them, and
submit them to the program.
The fuzzer collects information about the program
execution (\eg code coverage) to guide the next fuzz iteration.
Usually, fuzzers select seeds to improve the code coverage (code-coverage
guided fuzzers), but this behavior can be adjusted with different metrics and
purposes.

To select interesting seeds, fuzzers adopt two strategies: \emph{input
filtering} and \emph{queue culling}.
With input filtering, we refer to strategies that discard unproductive seeds,
while queue culling gives more priority to interesting seeds (without
discarding others).

Our work focuses on \emph{queue culling} strategies. Specifically, these
approaches use a specific flag, called \emph{favor}, to indicate whose seeds
will be selected in the next fuzz iteration.
The \emph{favor} setting can follow vary strategies according to the results
one wants to obtain. For instance, we can select seeds to improve the coverage,
or else we can guide the testing toward specific code locations in the attempt
to trigger a specific bug.

In case of AFL~\cite{afl}, it maintains a map (\texttt{top-rate}) that
pairs visited edges and the \emph{best} input for visiting it,
where the best input is simply the smallest and fastest to reach that edge.
While exercising inputs, AFL traces the visited edges and if the input results
better (\ie faster/smaller to reach it) for some edges in \texttt{top-rate},
than AFL assigns the new the input to those edges.
In this way, AFL can easily find suitable inputs for an edge through a fast
look-up.
More advanced fuzzers, such as Angora~\cite{chen2018angora} or
AFL-Sensitive~\cite{wang2019sensitive}, include additional information, such as
the calling stack, the memory access address, and the n-basic block execution
path.
More recent DGFs, such as TortoiseFuzz~\cite{wang2020not}, infer the best seed
based on a combination of static analysis and seed's execution path, \eg if the
seed has probability to hit sensitive code locations.
Regardless the complexity of the current cull queue algorithms, we observe a
main limitation: they do not consider dynamic information about the targets
status.
For instance, if a portion of code has been triggered, the cull queue should
stop considering that area as interesting and select seeds that hit other
locations. Conversely, current cull queue algorithms estimate the targets
priority ahead-of-time, without reconsidering them during the campaign.
To tackle such problems, \sys uses a novel \emph{queue culling} that employs a
dynamic target ranking that mutates during the fuzzing campaign
(\autoref{ssec:cull-queue}).

\subsection{Directed Greybox Fuzzers}
\label{ssec:directed-greybox-fuzzers}

Unlike traditional greybox fuzzers, which optimize for maximum code coverage,
DFG tries to reach specific code
locations~\cite{bohme2017directed}.
To achieve this goal, DFGs define distance-based mechanisms that direct the
fuzzers towards the targets via gradually reducing the
distance~\cite{bohme2017directed,chen2018hawkeye,osterlund2020parmesan}.
However, traditional solutions only calculate one harmonic distance between
seed and a set of targets, which try to cover all the target via one distance
and cannot scale to large number of targets.
Conversely, \sys proposes a dynamic multiple-distance measurement that
estimates the distance between the seeds and each target, respectively. Our
approach further expands the scope of the DGFs and scales them to large-scale
targets fuzzing.

\begin{figure}[t]
	\centering
	\includegraphics[width=0.6\linewidth]{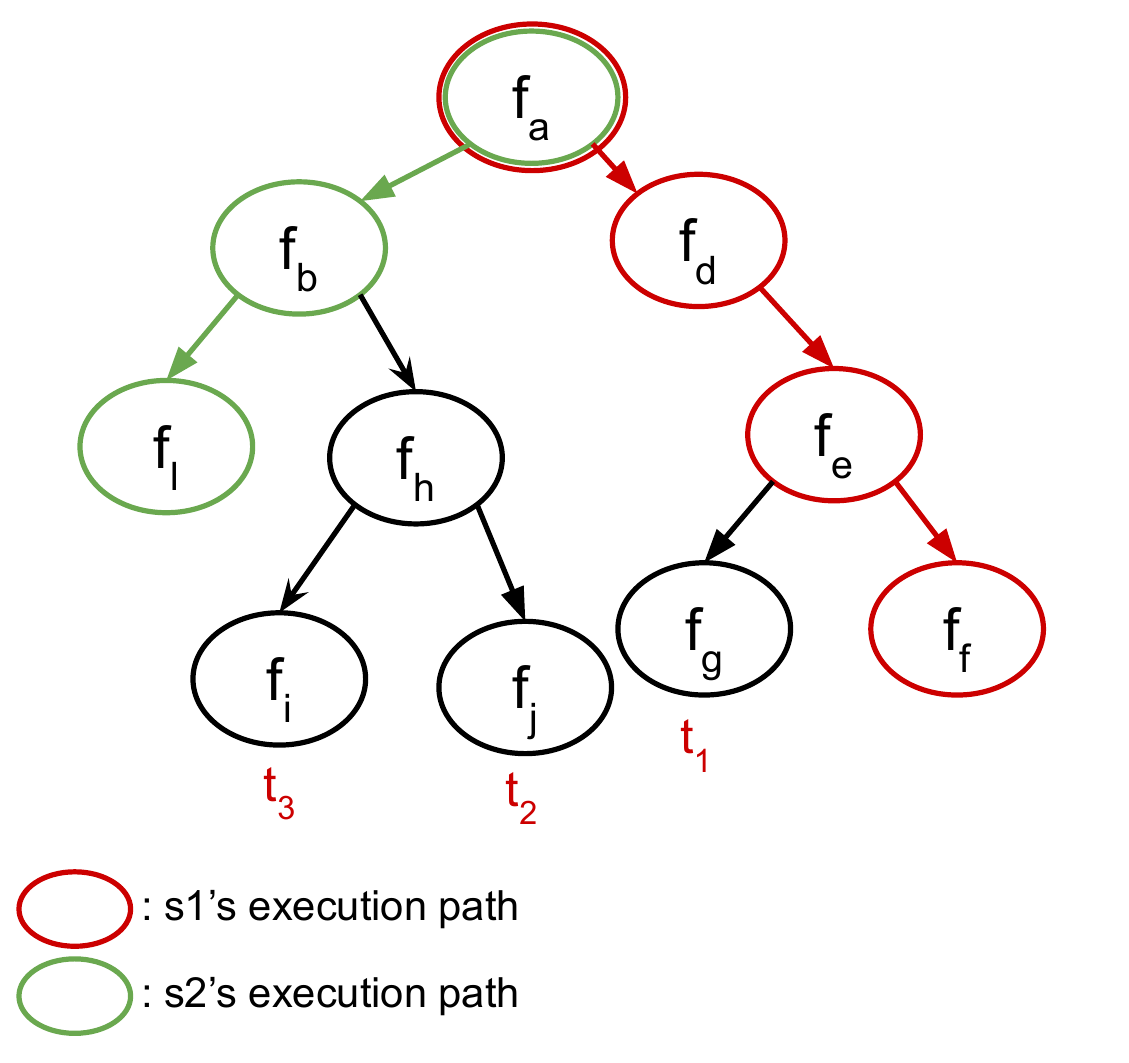}
	\caption{Example of seed-target distance for DGF.}
	\label{fig:multiple-distance}
\end{figure}

To show the limitations of current DGFs, we rely on the example in
\autoref{fig:multiple-distance}.
In this scenario, we assume having a call-graph in which we exercise two seeds,
$s_1$ and $s_2$, whose execution paths are red and green colored, respectively.
Moreover, we assume having a target set $T$ composed of $t_1$, $t_2$, and $t_3$.
Finally, we consider the graph's edges with an uniformly weighted as \emph{one}.
Without loss of generality, we apply to the graph a simplified version of the
harmonic-average distance used in AFLGo~\cite{bohme2017directed}
and ParmeSan~\cite{osterlund2020parmesan}. The estimation is done
in two steps. First, we compute the distance of each seed $s$ against each
target $t$ as the minimum number of edges between the execution path of $s$ and
the target $t$, \ie the distance between $s_1$ and $t_1$ is \emph{one} (the
edge $f_e$-$f_g$).
Then, we compute the harmonic-average among all the distances between the seeds
and the targets $T$, that result in $s_1 = 1.8$ and $s_2 = 2.25$. Having this
estimation, the fuzzer chooses $s_1$, thus privileging $t_1$.
The consequence is that the fuzzer becomes biased against $t_1$ and misjudge
$t_2$ and $t_3$. Since $t_2$ and $t_3$ fall far from $t_1$, the fuzzer hardly
mutates seeds in that direction.
Even worse, in case $t_1$ is triggered, the fuzzer will keep hitting $t_1$
since the distances only accounts graph's static information.
In \sys, we propose a novel multi-distance measure that mitigates such problems.


\section{\sys Design}

Directed fuzzers are hampered by two key problems:
\begin{itemize}
\item \textbf{P1: Seed explosion.} The fuzzing process produces a large amount
of seeds that are all selected in round robin fashion. This results in
promising seeds not receiving sufficient attention or, even worse, no fuzzing
cycles at all.
\item \textbf{P2: Imprecise distance estimation.} Calculating distance to many
targets (\eg all sanitizer labels) introduces imprecision and overwhelms the
selection prioritization algorithms of existing DGFs.
\end{itemize}

\sys introduces a smart seed selection strategy combined with a novel distance
metric to address P1 and P2.
Specifically, we implement the seed selection as a queue culling algorithm that
assigns the seed priority at each fuzzing iteration (more details in
\autoref{ssec:queue-culling-bkg}).
Our approach allows DGFs to handle programs with a large number of targets.

\sys dynamically adapts the priorities of the targets to privilege unexplored
ones---those that have not yet been covered and which are
more likely to contain unseen bugs, reducing the energy for those targets that
were already sufficiently explored.
Moreover, \sys  models the distance from seed to targets as a multi-distance function
that overcomes the limitations of previous works (\autoref{ssec:directed-greybox-fuzzers}).
Such mechanism works at function level and combine light static analyses with
dynamic information from the fuzzing session.

Our design is the result of two key limitations observed in the distance
estimation of previous works.
First, existing DGFs measure the distance between seeds and targets as a
single harmonic average. This approach synthesizes the information of all the
targets into a scalar~\cite{osterlund2020parmesan,bohme2017directed}
(\autoref{ssec:directed-greybox-fuzzers}).
While this works for a small set of few targets, it intuitively looses
precision when their number increases. For instance, the first work on
DGF dealt with tens of targets~\cite{bohme2017directed}, while \sys easily
scales to tens of thousands targets.
In other words, modeling the seed-targets distance with a single harmonic
average is equivalent to hit multiple targets with a single seed.
Secondly, all previous DGFs consider the targets as time-invariant, which means
the target importance is statically assigned before the campaign and cannot
change.
This intuitively dissipates fuzzing energy because we might end up hitting
unfruitful targets while overlooking more promising ones. For instance,
intensively tested targets are less likely to reveal new bugs, while poorly
explored targets can still express
errors~\cite{wagner2015high,osterlund2020parmesan}.
All these considerations drive the \sys seed selection strategy, as we
explain in the following sections.


\begin{figure*}[t]
	\centering
	\includegraphics[width=0.8\linewidth]{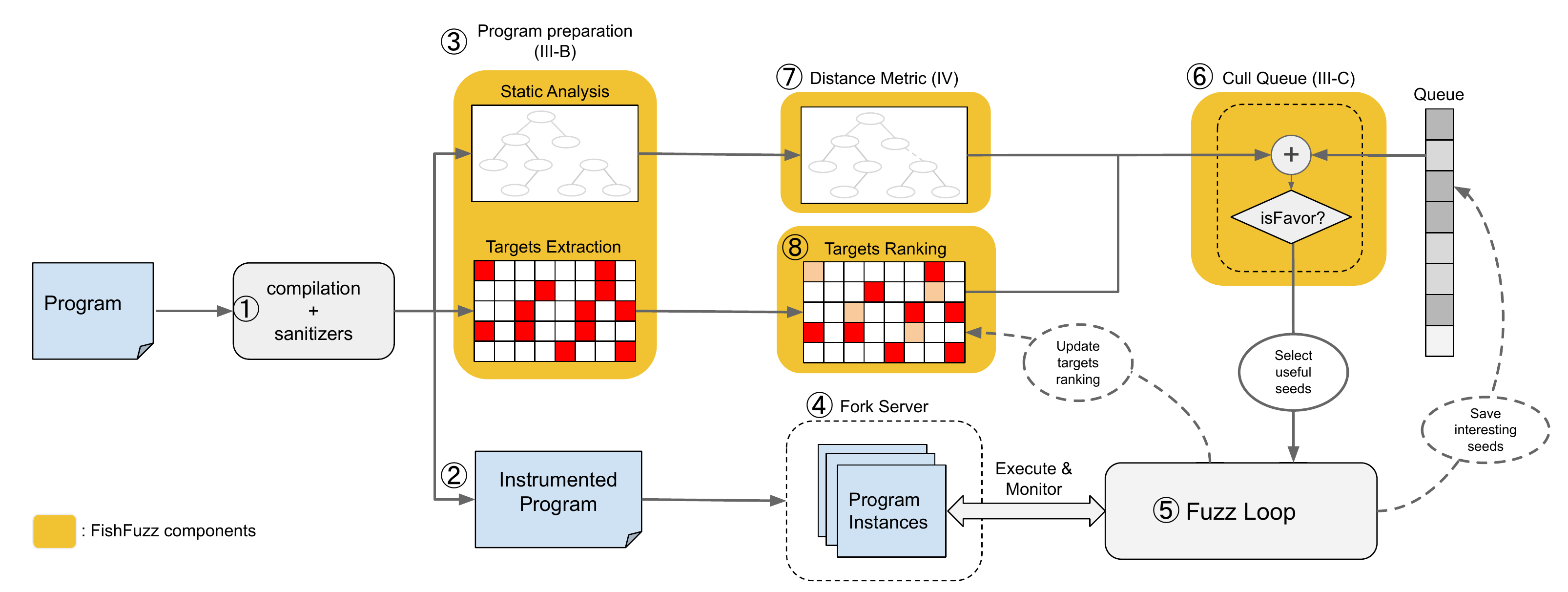}
	\caption{Overview of \sys workflow.}
	\label{fig:overview}
\end{figure*}

The design of \sys extends classic greybox
fuzzers~\cite{afl}. Our changes focus on a new queue culling
algorithm and a set of auxiliary structures to retain dynamic target information
and distance metrics.
Our design allows us to improve the seed selection based on extra
information not available otherwise.
The overall workflow is depicted in \autoref{fig:overview} and recalls
standard fuzzing procedures~\cite{afl,aflpp,wang2020not,osterlund2020parmesan}.

Given a target program, we compile and instrument it with specific sanitizers
(\circled[1]), this phase makes the instrumented program suitable for
the fuzzing campaign (\circled[2]), and extracts initial information
useful for \sys (\circled[3]).  We detail the latter in
\autoref{ssec:program-preprocessing}.
The instrumented program follows the standard greybox workflow in
which a fork server handles the program lifecycle (\circled[4]). Meanwhile, a
fuzz loop selects inputs from a queue and submits them to the program instances
(\circled[5]).
The input selection is handled by a queue culling algorithm (\circled[6]) that
relies on our novel distance metric (\circled[7]) and a dynamic target ranking
(\circled[8]). We detail the queue culling and the distance metric in
\autoref{ssec:cull-queue} and \autoref{sec:dyn-multitarget-distance},
respectively.
The target ranking is a shared structure that tracks meta information about the
targets, \eg hit frequency or if a target has been reached by a seed. We use
this information in the distance calculation and the cull queue.

\subsection{Program Preparation}
\label{ssec:program-preprocessing}

In the program analysis phase, we compile the program and generate a fuzzing compatibly binary.
We also instrument the code with extra components for code coverage
and security sanitizers similarly to previous
works~\cite{afl,osterlund2020parmesan,chen2018angora,wang2020not,bohme2017directed}.
The \sys design is agnostic by the sanitizer used, in our
experiment we successfully tested
ASan~\cite{asan,serebryany2012addresssanitizer} and UBSan~\cite{ubsan}.
We use the sanitizer information to extract the program targets.
Finally, we perform a lightweight static analysis at compilation time.

\paragraph*{Target Extraction}
Our fuzzer uses sanitizer check locations as target labels.
We rely on off-the-shelf sanitizers to extract
targets from the program.
Specifically, we locate the sanity checks injected on top of the original
program and consider them as targets to explore.
The intuition is that any sanity checks can potentially reveal the presence of
a bug. Our fuzzer then explores towards these code locations in the
attempt to trigger (exploit) them.
\sys is agnostic to the nature of the targets as long as they are uniquely
identifiable in the program.
As studied previously~\cite{osterlund2020parmesan,wagner2015high}, some
targets might be unreachable at runtime (\eg due to system environment).
Current approaches use static analysis to discard these cases, however, as the
same authors claim, using solely static analysis risks to also remove correct
sanity checks.
Conversely, \sys initially considers all the targets as valid and it deals with
false positive by dynamically ranking the targets and filtering out unpromising
ones.
Our approach removes possible errors from unsound static analysis (details in
\autoref{ssec:cull-queue}).

\paragraph*{Static Analysis}
During compilation, we extract the control-flow-graph (CFG) and the call-graph
(CG). This initial analysis, for now, is oblivious to indirect calls.
Then, \sys relies on CFG and CG in a novel inter-function distance to select
seeds closer to a given target.
We perform this operation at LLVM-IR~\cite{compiler-llvm}.
We describe the full function distances algorithm in
\autoref{ssec:function-distance} and discus how it deals with indirect calls in
\autoref{ssec:indirect-calls-handling}.

\subsection{Queue Culling Algorithm}
\label{ssec:cull-queue}

\begin{figure}[t]
	\centering
	\includegraphics[width=0.8\linewidth]{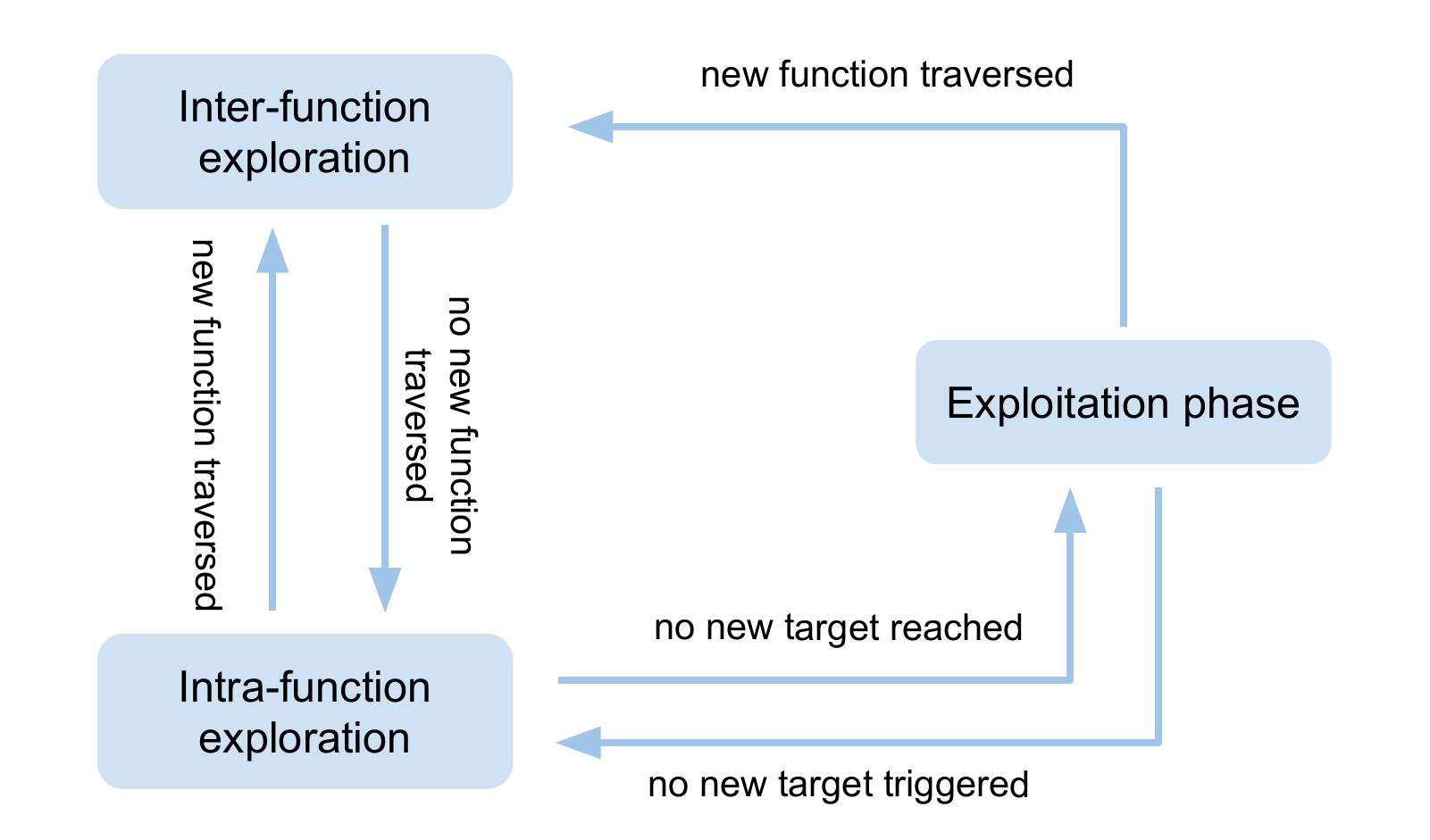}
	\caption{\sys uses an exploration phase to select seeds closer to
	targets, while the exploitation phase focuses on the targets triggering.
	Additionally, \sys uses two sub-exploration phases, one is specialized to
	reach larger number of functions (\emph{inter-function}), while the second
	to maximize the reached targets (\emph{intra-function}).}
	\label{fig:queue-fsm}
\end{figure}

We design the queue culling by taking inspiration from the \emph{trawl fishing}
technique. At the beginning of the fuzzing campaign, \sys prioritizes the
function exploration (expanding the net). When no new functions are reached,
\sys focuses on maximizing the reached targets (the net starts closing). Once
\emph{enough} targets are reached, the cull logic changes again and tries to
trigger the interesting targets (catching as many fish as possible).
Since every phase requires different metrics (\ie number of functions or
targets reached/triggered), we adopt a multiple phases approach as suggested
from previous works~\cite{osterlund2020parmesan}.
Specifically, \sys relies on three phases: \emph{inter-function
exploration}, \emph{intra-function} \emph{exploitation}, and
\emph{exploitation} -- all pictured \autoref{fig:queue-fsm}.
The purpose of the \emph{inter-function} exploration is to reach interesting
functions and it leverages our novel multi function-level distance (\ie
expanding the net). The \emph{intra-function} exploration, instead, focuses on
the internal function testing, relies on the standard AFL mutation algorithm, and tries
to hit as many targets as possible (\ie start closing the net).
Finally, the \emph{exploitation phase} drives the fuzzer energy to trigger
the maximum number of reachable targets (\ie catching the fish). This phase
uses a dynamic target ranking to prioritize promising locations.

The switch among the different phases happens at specific events:
\begin{enumerate*}[label=(\roman*)]
	\item every time \emph{new function is traversed},
	\item if \emph{no new function is found} for a period of time (\ie $30$min),
	\item if \emph{no new target is reached} for a period of time (\ie
	$10$min), and
	\item if \emph{no new target is triggered} for a period of time (\ie
	$1$hour).
\end{enumerate*}
In our experiments, we determined these timeouts for each event and leave
further per-target tuning as future work.

Our algorithm overcomes two shortcomings that affected previous works.
First, the \emph{inter-function} exploration uses a novel function distance
that is faster in selecting seeds closer to targets. This improves the
slow initial phase that affected previous
fuzzers~\cite{wang2020not,osterlund2020parmesan}.
Second, we boost the \emph{exploitation phase} with a multi-distance function
and a dynamic target ranking. Our approach discards non-profitable
targets, thus focusing on more likely bugs locations when comparing with the
state-of-the-art~\cite{wang2020not,osterlund2020parmesan,bohme2017directed}.

In the rest of this section, we detail the \emph{inter-function} exploration
and the \emph{exploitation} phase.
For what concerns the \emph{intra-function} exploration, we used the
standards AFL cull algorithm~\cite{afl}, we thus omit its description for
simplicity.

\paragraph*{Inter-function Exploration Phase}

\begin{algorithm}[t]
	\SetAlgoLined
	\DontPrintSemicolon
	\SetKwFunction{algo}{interFunctionCullQueue}
	\SetKwProg{myalg}{}{}{}
	\myalg{\algo{Queue, Functions}}{
		\For{$s \in$ Queue}{
			$s.favor = 0$\;
		}

		\For{$f \in$ Functions}{
			\If{$f.unexplored \wedge f.hastargets$}{ \label{alg:intercull-1}
				$s \gets getClosestSeedToFun(\emph{Queue},f)$\;
				$s.favor = 1$
			}
		}
	}
	\caption{Cull logic for the inter-function exploration phase.}
	\label{alg:interfunccull}
\end{algorithm}

In this phase, \sys selects seeds to maximize the reached functions
containing targets.
The cull algorithm of this phase is shown in \autoref{alg:interfunccull}.
Specifically, given a \emph{Queue} of seeds and a set of \emph{Functions} from
the target program, \sys sets $favor = 1$ to the closest seed for each
\emph{unexplored} function that also contains targets
(\autoref{alg:intercull-1}).
After the favored seeds are submitted to the program, \sys updates the list of
explored \emph{Functions} and repeat the process.
\emph{getClosestSeedToFun} finds the closest seed $s$ to the function $f$
through a seed-function distance that we discuss in
\autoref{ssec:function-distance}, in case of multiple seeds equally distant to
$f$, we prefer the lowest execution time.

\paragraph*{Exploitation Phase}

\begin{algorithm}[t]
	\SetAlgoLined
	\DontPrintSemicolon
	\SetKwFunction{algo}{exploitationCullQueue}
	\SetKwProg{myalg}{}{}{}
	\myalg{\algo{Queue, Targets}}{
		\For{$s \in$ Queue}{
			$s.favor = 0$\;
		}

		$trgs\_to\_visit \gets \emptyset$\;
		\For{$t \in$ Targets}{ \label{alg:exploitationcull-1}
			\If{$t.reached$}{
				$trgs\_to\_visit \gets trgs\_to\_visit \cup \{t\}$
			}
		}\label{alg:exploitationcull-2}

		$trgs\_to\_visit \gets orderByHit(trgs\_to\_visit)$\;
		$threshold \gets |trgs\_to\_visit | *
		20\%$\;\label{alg:exploitationcull-3}

		\For{$(p, t) \in enumerate(trgs\_to\_visit)$}{
			\If{$p < threshold$}{\label{alg:exploitationcull-4}
				$s \gets getFastestSeedToTarget(\emph{Queue}, t)$\;
				$s.favor = 1$
			}
		}
	}
	\caption{Cull logic for the exploitation phase.}
	\label{alg:exploitationcull}
\end{algorithm}

In the exploitation phase, \sys tries to trigger the maximum number of targets
previously reached. Our intuition is to keep hitting the same target with
different seeds (that can reach the target), thus increasing the chance to
expose a bug.
\autoref{alg:exploitationcull} shows the pseudo-code of this phase.
Specifically, we first select those targets that are reached through either the
\emph{inter-} or the \emph{intra-function} exploration phase
(\autoref{alg:exploitationcull-1} to \autoref{alg:exploitationcull-2}).
Among the \emph{trgs\_to\_visit}, we select the top 20\% of lesser hit targets
(\autoref{alg:exploitationcull-3} and \autoref{alg:exploitationcull-4}).
For each suitable target, \emph{getFastestSeedToTarget} returns the fastest
seed $s$ (with lowest execution time) that hits $t$, thus it finally sets it as
\emph{favor}.
In this phase we have high probability to have seeds that hit
targets (\ie seed-target distance $= 0$) due the exploration phase.
Finally, the function \emph{getFastestSeedToTarget} relies on a seed-target
multi-distance function that we detail in
\autoref{sec:dyn-multitarget-distance}.
In our prototype, we considered the top 20\% of lesser hit targets, we leave
the study of optimal threshold values as future work.

Our approach overcomes two important limitations of previous
DGFs~\cite{osterlund2020parmesan,wang2020not}. First, \sys has a dynamic view
of the target importance, \ie a triggered target looses importance
in the campaign, while less tested targets receive more energy.
Second, we automatically discard unreachable targets without employing heavy
software analysis~\cite{chen2018hawkeye,osterlund2020parmesan,wang2020not},
thus avoiding intrinsic false positives.
This philosophy is also reflected in the seed-target distance, as explained in
\autoref{sec:dyn-multitarget-distance}.

\section{Distance Measurement for \sys}
\label{sec:dyn-multitarget-distance}

\sys uses a novel function distance calculation that improves precision while
reducing complexity
compared to existing work. Specifically, our solution does not require heavy software
analysis to resolve indirect jumps.
Informally, this is done by mapping the distance between couple of functions,
\eg $\emph{dff}(f_a, f_b)$ indicates the distance between the function
$f_a$ and $f_b$.
\sys relies on this approach to either calculate seeds to function as well
as seeds to target distance.
The algorithm is composed on two steps. At compilation time, we analyze the
LLVM-IR code and build a \emph{static distance} map between functions
(\autoref{ssec:function-distance}).
In the fuzzing session, we leverage on the \emph{static distance} to estimate
the distance between seeds and a given function
(\autoref{sec:dynamic-seed-to-function-distance}).
We further rely on the \emph{static distance} to select the
closest seeds to a set of targets (\autoref{ssec:dyn-seed-to-multitarget-dist}).
In the last section, we discuss how our function distance deals with indirect
calls (\autoref{ssec:indirect-calls-handling}).

\subsection{Static Function Distance}
\label{ssec:function-distance}
Before the start of fuzzing, \sys generates a static map containing relationship
between functions.

\sys first assigns a \emph{weight} for each function pair $(f_i, f)$ such that
$f$ is a callee of $f_i$. The \emph{weight} represents the minimum number of
conditional edges that a seed might traverse from the entry point of $f_i$ to
the callee function $f$, and is computed with the function $dbb(m_a, m_b)$ (\ie
distance from basic block $m_a$ to $m_b$).
Formally speaking, given two functions $f_i$ and $f$, we defined $weight(f_i,
f)$ as follow:
\begin{equation} \label{eq:weight}
weight(f_i, f) =
\begin{cases}
\min {dbb(m, m_f)} & \text{if } \exists m_f \in f_i \\
\infty & \text{otherwise},
\end{cases}
\end{equation}
where $m$ is the first basic block of the function $f_i$, and $m_f$ is a basic
block belonging to $f_i$ and with a function call to function $f$ (so
$f$ is a callee of $f_i$).
If $f$ is a callee of $f_i$, the \emph{weight} between $f_i$ and $f$ is the
minimum distance between $m$ and $m_f$.
Otherwise, it is unreachable ($\infty$).
To handle multiple function calls to $f$, we consider only the minimum distance
to leave $f_i$.

Once the \emph{weights} are computed, \sys defines the distance between two
functions as the sum of their \emph{weight} along the shortest
path between two functions by following the CF extracted at compilation time.
Formally speaking, the distance between two function $f_a$ and $f_b$ is defined
as follow:
\begin{equation} \label{eq:dff}
\emph{dff}(f_a, f_b) = \sum_{f_i \in \emph{sp}(f_a, f_b)}{weight(f_i,
	f_{i+i})},
\end{equation}
where the function \emph{sp}($f_a, f_b$) returns the shortest path between
$f_a$ and $f_b$ using Dijkstra's algorithm~\cite{johnson1973note}, and
\emph{weight}($f_i, f_{i+1}$) is \autoref{eq:weight} over two consecutive
functions in the path.

\subsection{Dynamic Seed to Function Distance}
\label{sec:dynamic-seed-to-function-distance}
Having the static distance calculated in \autoref{ssec:function-distance}, we
define a function $\emph{dsf}(s, f)$ that represents the distance between the
functions traversed by the seed $s$ and a function $f$ as follow:
\begin{equation} \label{eq:dfs}
\emph{dsf}(s, f) =
\begin{cases}
\min_{f_s \in \xi(s)} \emph{dff}(f_s, f) & \text{if } f \not \in \xi(s) \\
0 & \text{otherwise},
\end{cases}
\end{equation}
where $\xi(s)$ is the set of functions traversed by the execution of the seed
$s$.
In case the $s$ already hits $f$, we consider the distance as \emph{zero}.

With the dynamic seed distance, \sys chooses the closest seed for a
target function upon the intuition that seeds closer to a target have
higher probability to reach it.
We mainly use \emph{dsf} in the inter-function exploration
(\autoref{ssec:cull-queue}).

%
%
%
%

\subsection{Dynamic Seed to Multi-Target Distance}
\label{ssec:dyn-seed-to-multitarget-dist}

\sys employs a novel multi-target distance to estimate seeds closer to a set of
targets $T_s$.
Differently from previous
works~\cite{bohme2017directed,osterlund2020parmesan,wang2020not}, which
represent the seed-targets distance a single harmonic average, \sys models the
seed-targets distance separately.
Precisely, \sys defines a function $D$ as follow:
\begin{equation}
D(s, T_s) \rightarrow (d_{t_1}, \dots, d_{t_x}),
\end{equation}
where $s$ is a seed, $T_s$ is a set of targets, and $d_{t_i}$ represents the
single distance between $s$ and the target $t_i$.
The size of the vector $(d_{t_1}, \dots, d_{t_x})$ is equal to the size of the
set $T_s$.

Each $d_{t_i}$ is defined as follow:
\begin{equation}
d_{t_i} = t_i.nottriggered~*~\emph{dfs}(s,f_{t_i}) \mid t_i \in f_{t_i},
\end{equation}
where $t_i.nottriggered$ is $1$ if $t_i$ has never been triggered, $0$
otherwise. \emph{dsf}($s,f_{t_i}$) is the function distance between the seed
$s$ and the function $f_{t_i}$ containing the target $t_i$ (\autoref{eq:dfs}).

This distance is used in the \emph{exploitation phase}
(\autoref{ssec:cull-queue}) and has two main advantages: first, it
automatically excludes targets already triggered (\ie $t_i.nottriggered$),
second, the distance between seed and targets is not affected by the size of
$T_s$.

\subsection{Indirect Call handling}
\label{ssec:indirect-calls-handling}

\begin{figure}[t]
	\centering
	\includegraphics[width=0.8\linewidth]{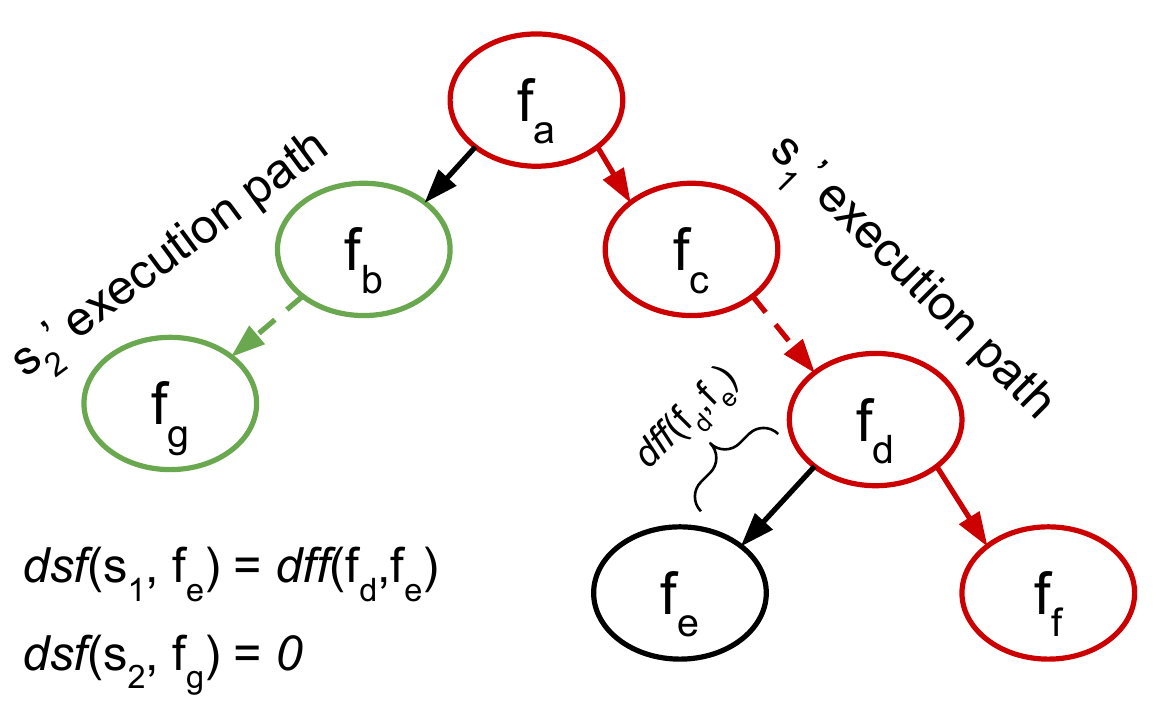}
	\caption{Example of indirect call handled by \sys. The picture shows a CF
	where $f_c$ and $f_d$, as well as $f_b$ and $b_g$, are connected through an
	unresolved indirect call.
	The execution path of the seed $s_1$ (red colored) traverses the indirect
	call between $F_c$ and $F_d$. This allows \sys to realize the existence of
	a path to $F_d$, and consequently to calculate the distance
	$\emph{dsf}(s_1, F_e) = \emph{dff}(F_d, F_e)$, which has been already
	extracted at compilation time.
	As a special case, a function could be isolate from the CG (\eg $f_g$).
	In this case, when a seed $s_2$ (green colored) this the function, we
	consider the distance as \emph{zero}.}
	\label{fig:indirect-call}
\end{figure}

Unresolved indirect calls might affect the quality of the fuzzing campaign.
Existing DGFs usually mitigate this issue by resolving the indirect jumps in
the CFG and CF.
For instance, Hawkeye~\cite{chen2018hawkeye} introduces inclusion-based pointer
analysis into CG generation, but this strategy does not cover all the indirect
calls. ParmeSan~\cite{osterlund2020parmesan} has an ad-hoc fuzzing session to
dynamically reconstruct the missing edges in  CFG and CG.

In \sys, the \emph{dynamic seed to function distance} (\autoref{eq:dfs})
already provides a good approximation of indirect calls without the burden to
resolve them. To explain this property, we rely on
\autoref{fig:indirect-call}, which shows a CF where the functions pair $f_c$ -
$f_d$, and $f_b$ - $f_g$, are connected through an (unresolved) indirect call.
This example explains two cases.
First, we assume the fuzzer has generated a seed $s_1$ that traverses
$f_a$, $f_c$, $f_d$, and finally hit $f_f$ (shown in red).
When this occurs, \sys has enough information to compute the distance
between $s$ and $f_e$, that is exactly the distance between $f_d$ and $f_e$.
Using our approach, we say we have an approximation of the distance because
we cannot estimate the component \emph{weight}($f_c$,$f_d$) (\autoref{eq:dff}).
However, since the distance seed-function is expressed as the \emph{minimal}
function distance, we argue our approach statistically finds a
\emph{quasi-optimal} result. The second case, instead, $f_g$ has no direct
calls to resolve the distance calculation. In this case, we assume the fuzzer
generates a seed $s_2$ that traverses $f_g$. According to \autoref{eq:dfs},
this results in distance \emph{zero} since $s_2$ hits the target function.
Generalizing the example, we can say that whenever a seed traverses an indirect
call, \sys can use near-by (connected) functions to estimate the minimum
distance between the seed's execution path and the target function.
In case a function has no direct connections, either the fuzzer generates a
seed that reaches the function or the latter is unreachable.
In practice, our approach is similar to previous works that use a fuzzing
session to explore indirect jumps~\cite{osterlund2020parmesan}. However, those
works apply such approach only once before fuzzing.
On the contrary, \sys benefits of each indirect jump resolved at any time of
the session.


\section{Implementation}
\label{sec:implementation}

The \sys implementation extends AFL~\cite{afl} version 2.57b and
LLVM~\cite{compiler-llvm} version 12.0.1\footnote{We also successfully test it
on LLVM 10.0.1}.

We implement the \emph{inter-function exploration} and the \emph{exploitation}
phases as two cull queue functions in AFL for a total of around $2,500$ LoC.
For the program analysis, we develop additionally analysis
passes for LLVM to extract CFG, CG, and estimate the \emph{static function
distance}. Moreover, we develop an additional instrumentation pass to extract
information for the \emph{dynamic seed to function} metric.
The LLVM code is around $1,500$ LoC in total.
Additionally, we have a few python scripts for the compilation process, which
is around $200$ LoC.
As for sanitizers, we use ASan~\cite{asan,serebryany2012addresssanitizer} and
UBSan~\cite{ubsan} distributed with the compiler-rt libraries from LLVM/Clang.

The source code of \sys, along with the material for replicating the
experiments, will be released open-source upon acceptance.

\section{Evaluation}
\label{sec:evaluation}

We evaluate the performance of \sys respect to the state of the art.
In particular, we desire to answer to the following research questions:
\begin{itemize}[leftmargin=*]
    \item[] \textbf{RQ1:} How many targets does \sys{} reach?
    (\autoref{ssec:exploration-study})
    \item[] \textbf{RQ2:} Does \sys{} balance the energy
    (\autoref{ssec:balance-energy})?
    \item[] \textbf{RQ3:} How efficiently does \sys{} find bugs
    (\autoref{ssec:speed-discovery-bugs})?
    \item[] \textbf{RQ4:} Can \sys{} find new bugs
    (\autoref{ssec:finding-new-bugs})?
    \item[] \textbf{RQ5:} How does \sys redistribute exploration and
    exploitation (\autoref{ssec:phases-distribution})?
    \item[] \textbf{RQ6:} Can other fuzzers benefit from our strategies 
    (\autoref{ssec:qsym-stuffs})?
\end{itemize}
All the experiments were exercised by following to the best practiced described
in~\cite{van_der_kouwe_sok_2019}.

\paragraph*{Comparison Works}
As comparison, we select three of the most modern and promising DGF in the
literature: TortoiseFuzz~\cite{wang2020not},
ParmeSan~\cite{osterlund2020parmesan}, and SAVIOR~\cite{chen2020savior}.
As a baseline, we choose AFL++~\cite{aflpp} and AFL~\cite{afl} as two of the 
most generic and coverage-based greybox fuzzer used in the community.
Moreover, we deploy \sys over QSYM~\cite{qsym} to answer to RQ6.
For our evaluation, we choose ASan and UBSan as sanitizers as mentioned in
\autoref{sec:implementation}.

\paragraph*{Experiment Setup}
All the experiments where performed on a Xeon Gold 5218 CPU (22M Cache, 2.30
GHz) equipped with 64GB of memory.
We evaluate all ASan targets on Ubuntu 22.04, but fall back to
Ubuntu 16.04 for the SAVIOR/QSYM + UBSan evaluation due to
compatibility issues of SAVIOR and QSYM  with newer versions of Ubuntu.
All experiments were run in docker containers with one core assigned.

\paragraph*{Benchmarks Selected}
We choose \numbenchmark benchmarks.
Specifically, two sets of programs come from
TortoiseFuzz~\cite{wang2020not} and SAVIOR~\cite{chen2020savior} that we
deploy over Ubuntu 22.04 and 16.04, respectively.
Since we use the TortoiseFuzz benchmark set with the ASan sanitizer, we call it 
the \emph{ASan benchmark}. Likewise, the SAVIOR benchmark contains only UBSan
sanitizers, thus we name it the \emph{UBSan benchmark}.
Regarding the ASan benchmark, we select \numasanprograms out of 10
programs and discard two of them due to incompatibility with Ubuntu
22.04.\footnote{\textsc{libming} and \textsc{catdoc} fail to compile on
Ubuntu 22.04 in their latest version.}
Additionally, only for ParmeSan, we remove $3$ programs from the ASan
benchmark due to an non-resolvable exception in ParmeSan---while the
other fuzzers handle all programs.
Regarding the UBSan benchmark, we choose \numubsanprograms and remove
one program because it does not compile on Ubuntu
16.04.\footnote{\textsc{objdump} runs out of memory during compilation.}
Finally, we compose a benchmark of \numrealworldprograms real programs for the
experiments in \autoref{ssec:finding-new-bugs}.



\subsection{RQ1: How many targets does \sys{} reach? }
\label{ssec:exploration-study}

We want to evaluate if the \emph{exploration phases} of \sys can reach more
targets respect similar DGFs.
To this end, we set two experiments, first, we exercise ParmeSan and
TortoiseFuzz against the ASan benchmark, then, SAVIOR against the UBSan
benchmark.
Finally, we evaluate \sys against both ASan adn UBSan benchmarks.
For ASan, we run $5$ rounds $60$ hours each, while for
UBSan, we run $5$ rounds $24$ hours each (as in the original
paper~\cite{chen2020savior,osterlund2020parmesan}).

The results for ASan and UBSan benchmarks are in \autoref{tab:ASanCov} and
\autoref{tab:UBSanCov}, respectively.
Likewise, we show the respective p-values of the Mann-Whitney U test in
\autoref{tab:ASanPVal} and \autoref{tab:UBSanPval}.

The figures show \sys reaches up to \bestasancovtort and \bestasancovparm more
edges  compared with TortoiseFuzz and ParmeSan on average, respectively.
Overall, \sys performs better than pure coverage-guided fuzzers like AFL++ by
reaching up to \bestasancovaflp more edges.
The only exceptional case was \textsc{exiv2}, for which \sys reached
\worstasancovaflp edges than AFL++ on average. This behavior can be explained
due to an over-optimization of the AFL++ instrumentation that allows to
exercise more seeds.\footnote{We are planning to port \sys over AFL++ for a
better comparison.}
Regardless a slightly drop in the \emph{exploration}, \sys manages to find more
unique bugs respect to AFL++ (more info in \autoref{ssec:speed-discovery-bugs}).
Furthermore, \sys shows an higher coverage in the UBSan benchmark, where we
reach up to \bestubsancovsav more edges compared with SAVIOR.

\sys expresses better performances also in terms of reached targets.
Specifically, we reach up to \bestasanreatort and \bestasanreaparm more targets
than TortoiseFuzz and ParmeSan, respectively.
Similarly for the coverage, \sys performs comparably with AFL++ by reaching up
to \bestasanreaaflp more targets. Again, here we notice a drop of
\worstasanreaaflp when compared with AFL++, this is expected since targets and
coverage are correlated measures.
Finally, we observe an improvement in the reached targets when comparing \sys
against SAVIOR of \bestubsanreasav on average at best.

\textbf{Takeaway:} Our experiment shows the \emph{exploration phase} of \sys 
can reach more targets respect to modern DGFs and with similar, if not better, 
results of AFL++.

\begin{table*}[t]
	\centering
	\caption{Edge coverage and targets reached in the ASan benchmark. The
	results refer to the average of $5$ rounds for $60$ hours each.}
	\label{tab:ASanCov}
	\resizebox{\textwidth}{!}{%
	\begin{tabular}{l|rr|rr|rr|rr|rr|rr|rr}
		\toprule
		\multirow{2}{*}{Program} &
		\multicolumn{2}{c|}{FishFuzz}                        &
		\multicolumn{2}{c|}{AFL++}                           &
		\multicolumn{2}{c|}{TortoiseFuzz}                    &
		\multicolumn{2}{c|}{ParmeSan}                        &
		\multicolumn{2}{c|}{vs AFL++}                        &
		\multicolumn{2}{c|}{vs TortoiseFuzz}                 &
		\multicolumn{2}{c}{vs ParmeSan}                      \\
		& \multicolumn{1}{c}{cov} & \multicolumn{1}{c|}{reach} &
		\multicolumn{1}{c}{cov} & \multicolumn{1}{c|}{reach} &
		\multicolumn{1}{c}{cov} & \multicolumn{1}{c|}{reach} &
		\multicolumn{1}{c}{cov} & \multicolumn{1}{c|}{reach} &
		\multicolumn{1}{c}{cov} & \multicolumn{1}{c|}{reach} &
		\multicolumn{1}{c}{cov} & \multicolumn{1}{c|}{reach} &
		\multicolumn{1}{c}{cov} & \multicolumn{1}{l}{reach}  \\ \midrule
		exiv2                    & 19484.2                 &
		9732.2                    & 20487.4                 &
		10125.0                   & 12290.6                 &
		7267.0                    & -                       &
		-                         & -4.90\%                 &
		-3.88\%                   & \textbf{58.53\%}        &
		\textbf{33.92\%}          & -                       &
		-                          \\
		flvmeta                  & 1010.0                  &
		153.0                     & 1009.2                  &
		153.0                     & 1008.2                  &
		153.0                     & 963                     &
		150                       & \textbf{0.08\%}         &
		0.00\%                    & \textbf{0.18\%}         &
		0.00\%                    & \textbf{4.92\%}         &
		\textbf{1.86\%}            \\
		gpac                     & 13910.4                 &
		1719.2                    & 9402.0                  &
		1226.0                    & 9638.8                  &
		1244.4                    & 5995                    &
		794                       & \textbf{47.95\%}        &
		\textbf{40.23\%}          & \textbf{44.32\%}        &
		\textbf{38.15\%}          & \textbf{132.02\%}       &
		\textbf{116.63\%}          \\
		liblouis                 & 2795.4                  &
		265.8                     & 2302.0                  &
		212.4                     & 2343.6                  &
		209.2                     & 1713                    &
		131                       & \textbf{21.43\%}        &
		\textbf{25.14\%}          & \textbf{19.28\%}        &
		\textbf{27.06\%}          & \textbf{63.21\%}        &
		\textbf{102.59\%}          \\
		libtiff                  & 16481.0                 &
		3021.8                    & 15741.6                 &
		2841.4                    & 14904.2                 &
		2678.4                    & 9290                    &
		1875                      & \textbf{4.70\%}         &
		\textbf{6.35\%}           & \textbf{10.58\%}        &
		\textbf{12.82\%}          & \textbf{77.41\%}        &
		\textbf{61.15\%}           \\
		nasm                     & 11709.0                 &
		1749.6                    & 11422.0                 &
		1666.4                    & 10659.8                 &
		1582.4                    & -                       &
		-                         & \textbf{2.51\%}         &
		\textbf{4.99\%}           & \textbf{9.84\%}         &
		\textbf{10.57\%}          & -                       &
		-                          \\
		ngiflib                  & 624.0                   &
		132.2                     & 615.0                   &
		123.6                     & 623.0                   &
		129.6                     & 459                     &
		86                        & \textbf{1.46\%}         &
		\textbf{6.96\%}           & \textbf{0.16\%}         &
		\textbf{2.01\%}           & \textbf{36.07\%}        &
		\textbf{53.72\%}           \\
		tcpreplay                & 1230.0                  &
		281.8                     & 1217.2                  &
		274.0                     & 1187.8                  &
		269.0                     & -                       &
		-                         & \textbf{1.05\%}         &
		\textbf{2.85\%}           & \textbf{3.55\%}         &
		\textbf{4.76\%}           & -                       &
		- \\
		\bottomrule
	\end{tabular}}
\end{table*}

\begin{table}[t]
	\centering
	\caption{Edge Coverage and Targets Reached in the UBSan benchmark. The
	results refer to the average of $5$ rounds for $24$ hours each.}
	\label{tab:UBSanCov}
	\resizebox{\columnwidth}{!}{%
	\begin{tabular}{l|rr|rr|rr}
		\toprule
		\multicolumn{1}{c|}{\multirow{2}{*}{Program}} &
		\multicolumn{2}{c|}{FishFuzz}                        &
		\multicolumn{2}{c|}{SAVIOR}                          &
		\multicolumn{2}{c}{vs SAVIOR}                        \\
		& \multicolumn{1}{c}{cov}
		& \multicolumn{1}{c|}{reach} & \multicolumn{1}{c}{cov} &
		\multicolumn{1}{c|}{reach} & \multicolumn{1}{c}{cov} &
		\multicolumn{1}{c}{reach}  \\ \midrule
		djpeg                                        & 12194.4
		& 3927.8                    & 11822.0                 &
		3863.8                    & \textbf{3.15\%}        &
		\textbf{1.66\%}           \\
		jasper                                       & 11126.4
		& 1823.0                    & 10878.4                 &
		1782.4                    & \textbf{2.28\%}        &
		\textbf{2.28\%}           \\
		readelf                                      & 2209.4
		& 181.8                     & 2101.8                  &
		170.8                     & \textbf{5.12\%}        &
		\textbf{6.44\%}           \\
		tcpdump                                      & 24316.4
		& 3747.0                    & 20041.8                 &
		3132.8                    & \textbf{21.33\%}       &
		\textbf{19.61\%}          \\
		tiff2pdf                                     & 13299.6
		& 1935.8                    & 12344.0                 &
		1744.6                    & \textbf{7.74\%}        &
		\textbf{10.96\%}          \\
		tiff2ps                                      & 9232.4
		& 1197.2                    & 8774.6                  &
		1141.8                    & \textbf{5.22\%}        &
		\textbf{4.85\%}           \\
		xmllint                                      & 8017.4
		& 723.8                     & 7855.2                  &
		713.8                     & \textbf{2.06\%}        &
		\textbf{1.40\%} \\
		\bottomrule
	\end{tabular}}
\end{table}

\begin{table*}[t]
	\centering
	\caption{p-values of the Mann-Whitney U test from the experiments in
	\autoref{tab:ASanCov}, while the column \textsc{uniq} refers to the unique
	bugs found.}
	\label{tab:ASanPVal}
	\begin{tabular}{l|rrr|rrr|rrr}
	\toprule
	\multirow{2}{*}{Program} &
	\multicolumn{3}{c|}{AFL++} &
	\multicolumn{3}{c|}{TortoiseFuzz} &
	\multicolumn{3}{c}{ParmeSan} \\
	& \multicolumn{1}{c}{cov} &
	\multicolumn{1}{c}{reach} & \multicolumn{1}{c|}{uniq} &
	\multicolumn{1}{c}{cov} & \multicolumn{1}{c}{reach} &
	\multicolumn{1}{c|}{uniq} & \multicolumn{1}{c}{cov} &
	\multicolumn{1}{c}{reach} & \multicolumn{1}{c}{uniq} \\ \midrule
	exiv2 & 0.2222 & 0.1508 &
	0.4884                    & 0.0119 &
	0.0119 & 0.1251                    &
	-      & -      &
	-                        \\
	flvmeta & 0.0200
	& 1.0000 & 1.0000                    &
	0.0056 & 1.0000 &
	1.0000                    & 0.0074 &
	0.0067 & 1.0000                   \\
	gpac & 0.0079
	& 0.0079 & 0.0114                    &
	0.0079 & 0.0079 &
	0.0114                    & 0.0079 &
	0.0079 & 0.0110                   \\
	liblouis                                      & 0.0317
	& 0.0159 & 0.0040                    &
	0.0159 & 0.0159 &
	0.0040                    & 0.0119 &
	0.0097 & 0.0040                   \\
	libtiff                                       & 0.0079
	& 0.0079 & 1.0000                    &
	0.0079 & 0.0079 &
	1.0000                    & 0.0079 &
	0.0079 & 1.0000                   \\
	nasm                                          & 0.0079
	& 0.0119 & 0.1770                    &
	0.0079 & 0.0119 &
	0.1770                    & -      &
	-      & -                        \\
	ngiflib                                       & 0.0926
	& 0.0937 & 1.0000                    &
	1.0000 & 0.2903 &
	1.0000                    & 0.0119 &
	0.0937 & 0.0040                   \\
	tcpreplay                                     & 0.4005
	& 0.3808 & 0.4065                    &
	0.0119 & 0.0092 &
	0.4237                    & -      &
	- & -\\ \bottomrule
	\end{tabular}%
\end{table*}

\begin{table}[]
	\centering
	\caption{p-values of the Mann-Whitney Y Test from the experiments in
	\autoref{tab:UBSanCov}, while the column \textsc{uniq} refers to the
	targets triggered.}
	\label{tab:UBSanPval}
	\begin{tabular}{l|lll}
	\toprule
	\multirow{2}{*}{Program} & \multicolumn{3}{c}{SAVIOR} \\
	 & \multicolumn{1}{c}{cov} & \multicolumn{1}{c}{reach} &
	 \multicolumn{1}{c}{uniq} \\ \midrule
	djpeg & 0.1508 & 0.2222 & 0.1798 \\
	jasper   & 0.0317 & 0.0079 & 0.0119 \\
	readelf  & 0.4019 & 0.4019 & 0.0111 \\
	tcpdump  & 0.0079 & 0.0079 & 0.0119 \\
	tiff2pdf & 0.0556 & 0.0556 & 0.0079 \\
	tiff2ps  & 0.0158 & 0.0556 & 0.0134 \\
	xmllint  & 0.0317 & 0.0749 & 0.0253 \\ \hline
	\end{tabular}
\end{table}

\begin{table}[t]
    \centering
    \caption{Unique bugs found in the ASan benchmark after $5$ round of $60$
    hour each. We report the best and the average round results.}
    \label{tab:ASanUniq}
    \begin{tabular}{l|rr|rr|rr|rr}
        \toprule
        \multirow{2}{*}{Program} &
        \multicolumn{2}{c|}{FishFuzz} &
         \multicolumn{2}{c|}{AFL++} &
        \multicolumn{2}{c|}{TortoiseFuzz} &
        \multicolumn{2}{c}{ParmeSan} \\
        & \multicolumn{1}{c}{best}      &
        \multicolumn{1}{c|}{avg}      &
        \multicolumn{1}{c}{best}      &
        \multicolumn{1}{c|}{avg}      &
        \multicolumn{1}{c}{best}      &
        \multicolumn{1}{c|}{avg}      &
        \multicolumn{1}{c}{best}      &
        \multicolumn{1}{c}{avg}       \\
        \midrule
        exiv2                    & \textbf{5}                           &
        \textbf{3.8}                         &
        4                                    &
        3.4                                  &
        4                                    &
        2.8                                  &
        -                                    &
        -                                     \\
        flvmeta                  & 0                                    &
        0.0                                  &
        0                                    &
        0.0                                  &
        0                                    &
        0.0                                  &
        0                                    &
        0.0                                   \\
        gpac                     & \textbf{22}                          &
        \textbf{16.0}                        &
        7                                    &
        5.6                                  &
        8                                    &
        6.2                                  &
        6                                    &
        6.0                                   \\
        liblouis                 & \textbf{3}                           &
        \textbf{3.0}                         &
        0                                    &
        0.0                                  &
        0                                    &
        0.0                                  &
        -                                    &
        -                                     \\
        libtiff                  & 1                                    &
        0.2                                  &
        1                                    &
        0.2                                  &
        1                                    &
        0.2                                  &
        0                                    &
        0.0                                   \\
        nasm                     & \textbf{1}                           &
        \textbf{0.4}                         &
        0                                    &
        0.0                                  &
        0                                    &
        0.0                                  &
        -                                    &
        -                                     \\
        ngiflib                  & 6                                    &
        6.0                                  &
        6                                    &
        6.0                                  &
        6                                    &
        6.0                                  &
        3                                    &
        3.0                                   \\
        tcpreplay                & \textbf{2}                           &
        \textbf{1.2}                         &
        2                                    &
        0.8                                  &
        1                                    &
        1.0                                  &
        -                                    &
        -                                     \\ \midrule
        sum                      & \textbf{40}                          &
        \textbf{30.6}                        &
        20                                   &
        16.0                                 &
        20                                   &
        16.2                                 &
        9                                    &
        9.0 \\
        \bottomrule
    \end{tabular}
\end{table}

\begin{table}[t]
	\centering
	\caption{Targets triggered in the UBSan benchmark upon $5$ rounds of $24$
	hours each. We report the best and the average round results.}
    \label{tab:UBSanUniq}
	\begin{tabular}{l|rr|rr|rr}
	\toprule
	\multirow{2}{*}{Program}  & \multicolumn{2}{c|}{FishFuzz} &
	\multicolumn{2}{c|}{SAVIOR} &
	\multicolumn{2}{c}{vs SAVIOR} \\
	& \multicolumn{1}{c}{best} & \multicolumn{1}{c|}{avg} &
	\multicolumn{1}{c}{best} & \multicolumn{1}{c|}{avg} &
	\multicolumn{1}{c}{best} & \multicolumn{1}{c}{avg} \\ \midrule
	djpeg & 145  & 138 & 134  & 134 & \textbf{+8.21\%} & \textbf{+2.99\%}  \\
	jasper & 54 & 50.8 & 47 & 41.4 & \textbf{+14.89\%} & \textbf{+22.71\%} \\
	readelf  & 39 & 31.4 & 24 & 21.6 & \textbf{+62.50\%} & \textbf{+45.37\%} \\
	tcpdump  & 169 & 143.8 & 110 & 107.4 & \textbf{+53.64\%} &
	\textbf{+33.89\%} \\
	tiff2pdf & 23 & 21 & 17 & 13.8 & \textbf{+35.29\%} & \textbf{+52.17\%} \\
	tiff2ps  & 17 & 15.2 & 14 & 12.4 & \textbf{+21.43\%} & \textbf{+22.58\%} \\
	xmllint  & 17 & 15.4 & 14 & 13.8 & \textbf{+21.43\%} & \textbf{+11.59\%} \\
	\bottomrule
	\end{tabular}%
\end{table}

\begin{table*}[]
	\centering
	\caption{Time-To-Exposure of $47$ real world bugs found after $5$ rounds of
	$60$ hours in the ASan benchmark.}
    \label{tab:ASanReal}
    \resizebox{\textwidth}{!}{%
	\begin{tabular}{l|l|l|rr|rr|rr|rr}
	\toprule
	\multirow{2}{*}{Program} & \multirow{2}{*}{Location} &
	\multicolumn{1}{c|}{\multirow{2}{*}{Bug}} &
	\multicolumn{2}{c|}{FishFuzz} & \multicolumn{2}{c|}{AFL++} &
	\multicolumn{2}{c|}{TortoiseFuzz} & \multicolumn{2}{c}{ParmeSan} \\
	& & & \multicolumn{1}{c}{best} & \multicolumn{1}{c|}{avg} &
	\multicolumn{1}{c}{best} & \multicolumn{1}{c|}{avg} &
	\multicolumn{1}{c}{best} & \multicolumn{1}{c|}{avg} &
	\multicolumn{1}{c}{best} & \multicolumn{1}{c}{avg} \\ \midrule
	exiv2 &  decodeIHDRChunk-\textgreater{}getLong & issue\_170 &
	\textless{}1m & \textless{}1m & \textless{}1m & \textless{}1m &
	\textless{}1m & \textless{}1m & - & - \\
	exiv2           &
	decodeTXTChunk-\textgreater{}keyTXTChunk
	       &
	 CVE-2017-17669           &
	\textless{}1m &
	\textless{}1m                                  &
	\textless{}1m                          &
	\textless{}1m                          &
	\textless{}1m                          &
	\textless{}1m                          &
	- & -\\
	exiv2           &
	printTiffStructure-\textgreater{}printIFDStructure
	       &
	 CVE-2017-12955           & \textbf{23.17m}
	& 7.54h                                          &
	1.29h                                  & \textbf{7.31h}  &
	3.41h                                  &
	38.37h                                 &
	-                                             &
	-                                             \\
	exiv2           &
	printIFDStructure-\textgreater{}printIFDStructure
	       &
	 CVE-2017-14861           & \textbf{3.03h}
	&  \textbf{28.14h}         &
	9.43h                                  &
	41.48h                                 &
	8.53h                                  &
	49.71h                                 &
	-                                             &
	-                                             \\
	exiv2           &
	doAccept-\textgreater{}visitDirectory
	       &
	 unknown-1                & \textbf{49.83h}
	&  \textbf{57.97h}         &
	60h                                   &
	60h                                    &
	60h                                    &
	60h                                    &
	-                                             &
	-                                             \\
	MP4Box          & gf\_isom\_box\_array\_dump
	-\textgreater{}gf\_isom\_box\_dump\_ex                  &
	CVE-2019-20168           & \textless{}1m                                 &
	 \textbf{\textless{}1 m} &
	\textless{}1m                          &
	1.27m                                  &
	\textless{}1m                          &
	2.10m                                  &
	3.00m                                         &
	13.18h                                        \\
	MP4Box          & gf\_isom\_box\_del
	-\textgreater{}gf\_isom\_box\_del                               &
	CVE-2020-11558           & 1.58m                                         &
	1.83m                                          &
	1.12m                                  &
	1.23m                                  &
	8.02m                                  &
	8.11m                                  &
	\textbf{\textless{}1m} &  \textbf{\textless{}1m} \\
	MP4Box          & gf\_import\_mpeg\_ts
	-\textgreater{}gf\_m2ts\_process\_data                        &
	CVE-2019-13618           & \textbf{1.03h}         &
	\textbf{12.77h}         &
	22.02h                                 &
	45.20h                                 &
	20.89h                                 &
	42.60h                                 &
	60h                                           &
	60h                                           \\

	MP4Box          & gf\_m2ts\_gather\_section
	-\textgreater{}gf\_m2ts\_section\_complete               &
	CVE-2020-24829           & \textbf{1.33h}       &
	 \textbf{2.13h}          &
	46.38h                                 &
	57.28h                                 &
	60h                                    &
	60h                                    &
	60h                                           &
	60h                                           \\
	MP4Box          & gf\_isom\_box\_parse\_ex
	-\textgreater{}gf\_isom\_oinf\_read\_entry                &
	CVE-2019-20169           & 1.52h                                         &
	2.59h                                          &
	\textbf{14.05m} &  \textbf{41.76m} &
	33.66m                                 &
	1.12h                                  &
	6.00m                                         &
	57.44m                                        \\
	MP4Box          & gf\_m2ts\_process\_data
	-\textgreater{}gf\_m2ts\_get\_adaptation\_field            &
	issue\_1446              & \textbf{1.77h}         &
	 \textbf{3.28h}          &
	51.87h                                 &
	58.37h                                 &
	60h                                    &
	60h                                    &
	60h                                           &
	60h                                           \\
	MP4Box          & gf\_isom\_box\_parse\_ex
	-\textgreater{}urn\_Read                                  &
	CVE-2018-13005           & \textbf{3.89h}         &
	 \textbf{7.59h}          &
	29.50h                                 &
	53.31h                                 &
	6.30h                                  &
	32.23h                                 &
	37.31h                                        &
	55.46h                                        \\
	MP4Box          & hdlr\_dump
	-\textgreater{}\_\_interceptor\_strlen.part.0                           &
	CVE-2018-13006           & \textbf{12.23h}        &
	 \textbf{17.35h}         &
	60h                                    &
	60h                                    &
	39.52h                                 &
	55.90h                                 &
	25.35h                                        &
	53.07h                                        \\
	MP4Box          & gf\_bs\_read\_int
	-\textgreater{}BS\_ReadByte                                      &
	unknown-2                & \textbf{4.66h}         &
	 \textbf{26.25h}         &
	60h                                    &
	60h                                    &
	60h                                    &
	60h                                    &
	60h                                           &
	60h                                           \\
	MP4Box          & gf\_bs\_read\_data
	-\textgreater{}\_\_asan\_memcpy                                 &
	unknown-3                & 9.43h                                         &
	36.07h                                         &
	60h                                    &
	60h                                    &
	60h                                    &
	60h                                    &
	\textbf{26.00m}        &  \textbf{3.28h}         \\
	MP4Box          & gf\_m2ts\_section\_complete
	-\textgreater{}gf\_m2ts\_process\_sdt                  &
	issue\_1426              & \textbf{15.21h}        &
	 \textbf{26.24h}         &
	60h                                    &
	60h                                    &
	60h                                    &
	60h                                    &
	60h                                           &
	60h                                           \\
	MP4Box          & gf\_m2ts\_section\_complete
	-\textgreater{}gf\_m2ts\_process\_pmt                  &
	issue\_1421              & \textbf{3.79h}         &
	 \textbf{37.81h}         &
	60h                                    &
	60h                                    &
	60h                                    &
	60h                                    &
	60h                                           &
	60h                                          \\
	MP4Box          & avcc\_Read
	-\textgreater{}gf\_media\_avc\_read\_sps                                &
	CVE-2020-22678           & 19.58h                                        &
	44.43h                                         &
	\textbf{1.99h}  & \textbf{7.97h}  &
	30.20h                                 &
	50.18h                                 &
	4.07h                                         &
	32.07h                                        \\
	MP4Box          & gf\_isom\_open\_file
	-\textgreater{}gf\_isom\_parse\_movie\_boxes                  &
	issue\_2092              & 60h                                           &
	60h                                            &
	\textbf{52.25h} & \textbf{58.45h} &
	60h                                    &
	60h                                    &
	60h                                           &
	60h                                           \\
	MP4Box          & AVC\_RewriteESDescriptorEx
	-\textgreater{}gf\_isom\_sample\_entry\_get\_bitrate    &
	issue\_1180              & 20.58h                                        &
	44.82h                                         &
	54.06h                                 &
	58.81h                                 &
	\textbf{7.31h}  &  \textbf{34.03h} &
	60h                                           &
	60h                                           \\
	MP4Box          & gf\_isom\_box\_array\_read\_ex
	-\textgreater{}stbl\_AddBox                         &
	issue\_1332              & 60h                                           &
	60h                                            &
	60h                                    &
	60h                                    &
	60h                                    &
	60h                                    &
	\textbf{8.22h}         &  \textbf{39.69h}       \\
	MP4Box          & HEVC\_RewriteESDescriptorEx
	-\textgreater{}gf\_isom\_sample\_entry\_get\_bitrate &
	unknown-4                & \textbf{14.98h}        &
	\textbf{39.92h} & 60h &	60h & 60h &	60h
	&
	60h & 60h \\
	MP4Box          & gf\_odf\_delete\_descriptor-\textgreater{}free
	-\textgreater{}asan\_free           & issue\_1271              &
	\textbf{9.83h}         & \textbf{43.28h}         &
	60h                                    &
	60h                                    &
	60h                                    &
	60h                                    &
	60h                                           &
	60h                                           \\
	MP4Box          & gf\_odf\_delete\_descriptor\_list
	-\textgreater{}gf\_list\_enum                    & unknown-5
	& \textbf{10.00h}       & \textbf{43.37h}         &
	60h                                    &
	60h                                    &
	60h                                    &
	60h                                    &
	60h                                           &
	60h                                           \\
	MP4Box          &
	gf\_odf\_delete\_descriptor-\textgreater{}gf\_odf\_delete\_descriptor
	       & unknown-6                &
	\textbf{10.34h}        &  \textbf{43.44h}         &
	60h                                    &
	60h                                    &
	60h                                    &
	60h                                    &
	60h                                           &
	60h                                           \\
	MP4Box          & gf\_ipmpx\_data\_parse
	-\textgreater{}GF\_IPMPX\_ReadData                          &
	issue\_1320              & \textbf{27.67h}        &
	 \textbf{53.53h}         &
	60h                                    &
	60h                                    &
	60h                                    &
	60h                                    &
	60h                                           &
	60h                                           \\
	MP4Box          & gf\_ipmpx\_data\_del
	-\textgreater{}GF\_IPMPX\_AUTH\_Delete                        &
	issue\_1328              & \textbf{36.25h}       &
	 \textbf{55.25h}         &
	60h                                   &
	60h                                    &
	60h                                    &
	60h                                    &
	60h                                           &
	60h                                           \\
	MP4Box          & gf\_m2ts\_process\_pmt
	-\textgreater{}on\_m2ts\_import\_data                       &
	issue\_1446              & \textbf{34.30h  }        &
	 \textbf{50.27h}         &
	60h                                    &
	60h                                    &
	60h                                    &
	60h                                    &
	60h                                           &
	60h                                           \\
	MP4Box          & AVC\_RewriteESDescriptorEx
	-\textgreater{}AVC\_DuplicateConfig                     &
	issue\_1179              & \textbf{26.12h}        &
	 \textbf{50.54h}         &
	60h                                    &
	60h                                    &
	60h                                    &
	60h                                    &
	60h                                           &
	60h                                           \\
	MP4Box          & gf\_isom\_box\_parse\_ex
	-\textgreater{}dimC\_Read                                 &
	issue\_1348              & \textbf{28.18h}        &
	 \textbf{53.64h}         &
	60h                                    &
	60h                                    &
	60h                                    &
	60h                                    &
	60h                                           &
	60h                                           \\
	MP4Box          & gf\_m2ts\_reframe\_ac3
	-\textgreater{}gf\_ac3\_parser                              &
	unknown-7                & \textbf{38.19h }        &
	 \textbf{55.64h}         &
	60h                                    &
	60h                                    &
	60h                                    &
	60h                                    &
	60h                                           &
	60h                                           \\
	MP4Box          & gf\_m2ts\_reframe\_mpeg\_audio
	-\textgreater{}gf\_mp3\_get\_next\_header\_mem      &
 	unknown-8                & \textbf{41.11h}            &
	 \textbf{56.22h}         &
	60h                                    &
	60h                                    &
	60h                                    &
	60h                                    &
	60h                                           &
	60h                                           \\
	MP4Box          & gf\_m2ts\_reframe\_nalu\_video
	-\textgreater{}\_\_interceptor\_memchr.part.0       &
    	unknown-9                & \textbf{41.50h}        &
	 \textbf{56.30h}         &
	60h                                    &
	60h                                    &
	60h                                    &
	60h                                    &
	60h                                           &
	60h                                           \\
	MP4Box          & gf\_m2ts\_flush\_pes
	-\textgreater{}gf\_m2ts\_reframe\_aac\_adts                   &
	unknown-10               & \textbf{41.75h}        &
	 \textbf{56.35h}         &
	60h                                    &
	60h                                    &
	60h                                    &
	60h                                    &
	60h                                           &
	60h                                           \\
	lou\_checktable & \_\_interceptor\_vsnprintf
	-\textgreater{}printf\_common                           &
	issue\_728               & \textbf{10.33h}        &
	 \textbf{23.27h}         &
	60h                                    &
	60h                                    &
	60h                                    &
	60h                                    &
	60h                                           &
	60h                                           \\
	lou\_checktable & compileRule
	-\textgreater{}compileUplow                                            &
	CVE-2018-11410           & \textbf{16.72h}        &
	 \textbf{39.94h}         &
	60h                                    &
	60h                                    &
	60h                                    &
	60h                                    &
	60h                                           &
	60h                                           \\
	lou\_checktable & pattern\_compile\_expression-\textgreater
	pattern\_compile\_expression             & issue\_573               &
	\textbf{35.43h}        &
	\textbf{42.26h}         & 60h                                    &
	60h                                    &
	60h                                    &
	60h                                    &
	60h                                           &
	60h                                           \\
	gif2tga         & DecodeGifImg
	-\textgreater{}WritePixel                                             &
	issue\_11                & \textless{}1m                                 &
	\textless{}1m                                  &
	\textless{}1m                          &
	\textless{}1m                          &
	\textless{}1m                          &
	\textless{}1m                          &
	60h                                           &
	60h                                           \\
	gif2tga         & DecodeGifImg
	-\textgreater{}WritePixels                                            &
	issue\_1                 & \textless{}1m                                 &
	\textless{}1m                                  &
	\textless{}1m                          &
	\textless{}1m                          &
	\textless{}1m                          &
	\textless{}1m                          &
	\textless{}1m                                 &
	\textless{}1m                                 \\
	gif2tga         & WritePixels
	-\textgreater{}GifIndexToTrueColor                                     &
	CVE-2019-20219           & \textless{}1m                                 &
	\textless{}1m                                  &
	\textless{}1m                          &
	\textless{}1m                          &
	\textless{}1m                          &
	\textless{}1m                          &
	60h                                           &
	60h                                           \\
	gif2tga         & LoadGif
	-\textgreater{}DecodeGifImg                                                &
	 issue\_4                 & \textless{}1m
	& \textless{}1m                                  &
	\textless{}1m                          &
	\textless{}1m                          &
	\textless{}1m                          &
	\textless{}1m                          &
	\textless{}1m                                 &
	2.30m                                         \\
	gif2tga         & WritePixel
	-\textgreater{}GifIndexToTrueColor                                      &
	issue\_9-2               & \textless{}1m                                 &
	 \textbf{\textless{}1m}  &
	\textless{}1m                          &
	2.08m                                  &
	1.41m                                  &
	2.36m                                  &
	\textless{}1m                                 &
	5.00m                                         \\
	gif2tga         & FillGifBackGround
	-\textgreater{}GifIndexToTrueColor                               &
	issue\_14                & \textbf{\textless{}1m} &
	 \textbf{1.44m}          &
	1.62m                                  &
	2.61m                                  &
	1.62m                                  &
	5.25m                                  &
	60h                                           &
	60h                                           \\
	tcpprep         & get\_ipv4
	-\textgreater{}get\_l2len\_protocol                                      &
	issue\_716               & \textbf{\textless{}1m} &
	 \textbf{\textless{}1m}  &
	11.46m                                 &
	32.80h                                 &
	2.37h                                  &
	8.68h                                  &
	-                                             &
	-                                             \\
	tcpprep         & parse\_metadata
	-\textgreater{}parse\_mpls                                         &
	issue\_719               & \textbf{14.09h}         &
	 \textbf{50.82h}         &
	40.07h                                 &
	56.01h                                 &
	60h                                    &
	60h  & - &   - \\
	tiff2pdf        & t2p\_write\_pdf
	-\textgreater{}t2p\_read\_tiff\_data                               &
	CVE-2017-9935            & 34.66h                                        &
	54.93h                                         &
	59.11h                                 &
	59.82h                                 &
	\textbf{31.13h} &  \textbf{54.23h} &
	60h & 60h \\
	nasm            & scan
	-\textgreater{}ppscan
	  &
	 CVE-2019-6291            & \textbf{11.22h}
	& \textbf{42.86h}         &
	60h                                    &
	60h                                    &
	60h                                    &
	60h                                    &
	-                                             &
	-                                             \\ \bottomrule
	\end{tabular}%
    }
\end{table*}

\subsection{RQ2: Does \sys{} balance the energy?}
\label{ssec:balance-energy}

We asses the ability of \sys to redistribute energy among the targets. 
For this evaluation, we choose AFL as baseline because it is the base code for
our prototype, therefore, we can better appreciate the improvements from our
methodology.
For the experiment, we run \sys and AFL against the UBSan
benchmark for $3$ rounds of $24$ hours each. Then, we average the target visit 
frequency. Finally, we order the targets by visit frequency and plot them in
\autoref{fig:energy_balanced}.

The combination of the \sys culling algorithm (\autoref{ssec:cull-queue}) and
the target distance (\autoref{ssec:dyn-seed-to-multitarget-dist}) tend to
re-assign energy to the lesser tested targets.
This is reflected in \autoref{fig:energy_balanced} where \sys shows fewer
targets with \emph{zero} frequency in the distribution tails.
Specifically, this is evident for $5$ out of $7$ programs in which \sys
expresses a better balanced energy respect to AFL (\ie all targets have been
visited at least once).
For \textsc{tcpdump}, we observe \sys has a few non-visited targets,
but overall the curve is better redistributed.
\textsc{jasper} is the only case where AFL seems to have a slightly
better balancing. We further investigate \textsc{jasper} and notice this
behavior is caused by \sys that discovers new targets in the last part of the
campaign.
Consequently, \sys has not time to assign energy to them.

\textbf{Takeaway:} Overall, we show that \sys can effectively redistribute the
energy compared with a fair baseline. This leads to a better target
exploitation phase and to statistically increase the chance to find new bugs.

\newcommand{\subimagewidth}{0.23\textwidth}
\begin{figure}[t]
    \centering
    \begin{tabular}{p{\subimagewidth} p{\subimagewidth}}

        \includegraphics[width=\subimagewidth]{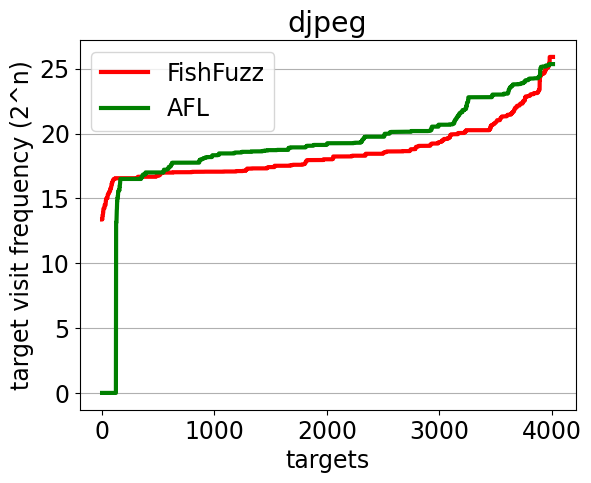} &
        \includegraphics[width=\subimagewidth]{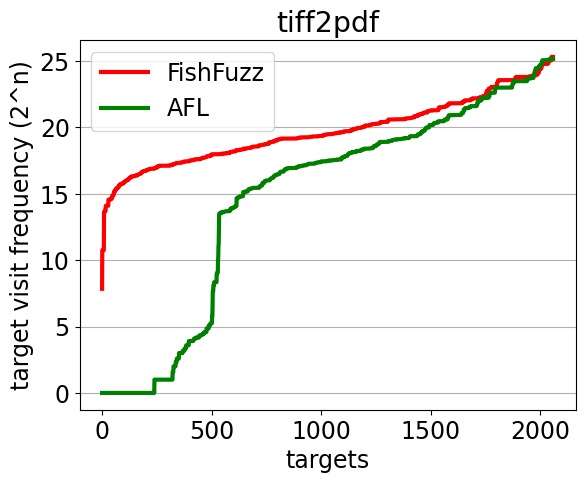} \\
        \includegraphics[width=\subimagewidth]{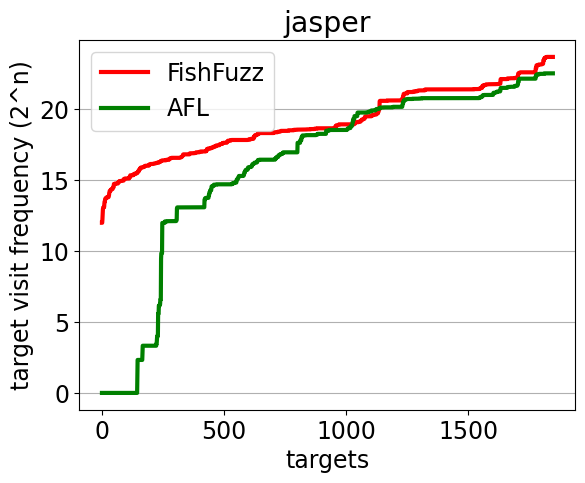} &
        \includegraphics[width=\subimagewidth]{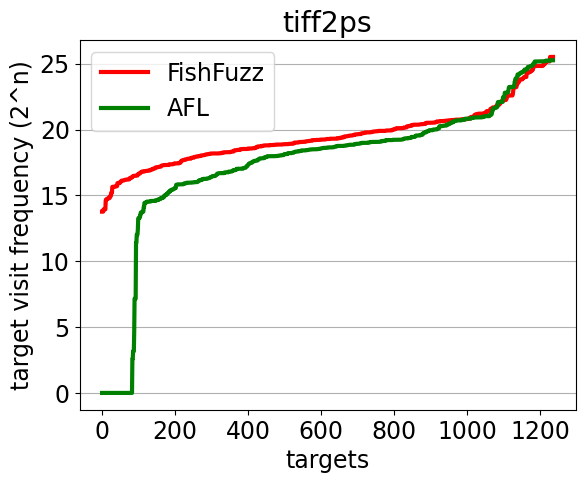} \\
        \includegraphics[width=\subimagewidth]{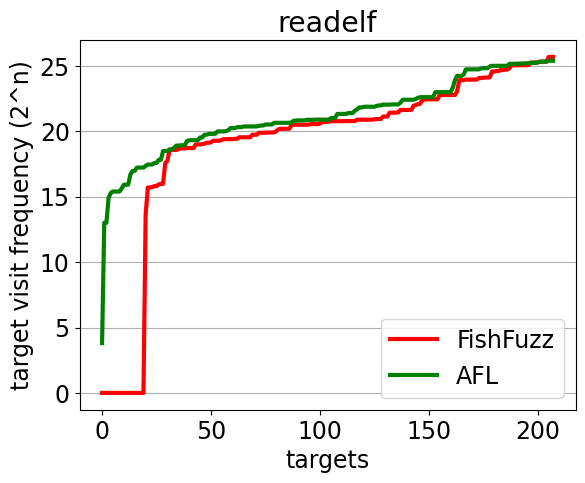} &
        \includegraphics[width=\subimagewidth]{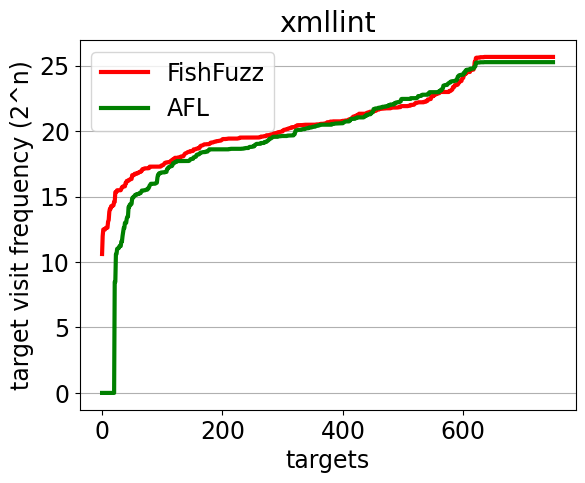} \\

        \multicolumn{2}{c}{\includegraphics[width=\subimagewidth]{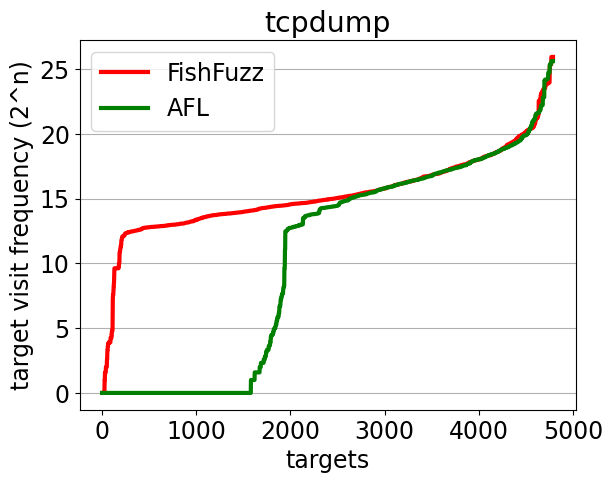}}
        \\

    \end{tabular}
    \caption{Energy allocated by \sys{} compared against AFL as baseline.
    \sys re-assigns energy to lesser explored targets, thus resulting in a tail
    with lesser non-tested targets. For \textsc{jasper}, we observe a few
    targets with \emph{zero} visits because discovered in the latest phase of
    the campaign.}
    \label{fig:energy_balanced}
\end{figure}

\subsection{RQ3: How efficiently does \sys{} find bugs?}
\label{ssec:speed-discovery-bugs}
%

\begin{figure}[t]
	\centering
		\includegraphics[width=0.25\textwidth]{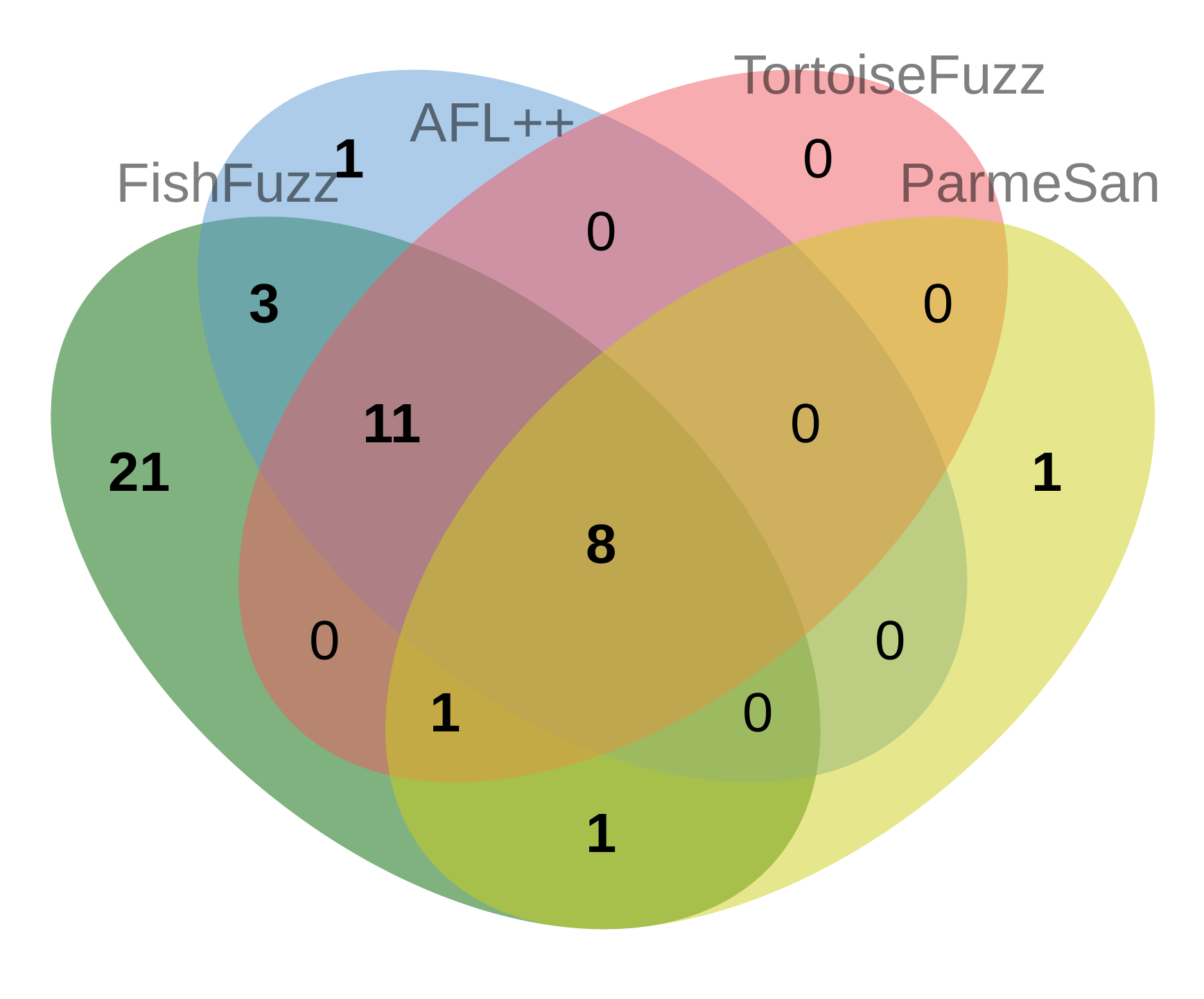}
		\caption{Unique bugs found in the ASan benchmark and compared with 
			\sys, AFL++, TortoiseFuzz, and ParmeSan.}
		\label{fig:venn_asan}
\end{figure}

\begin{figure}[t]
	\centering
	\includegraphics[width=0.25\textwidth]{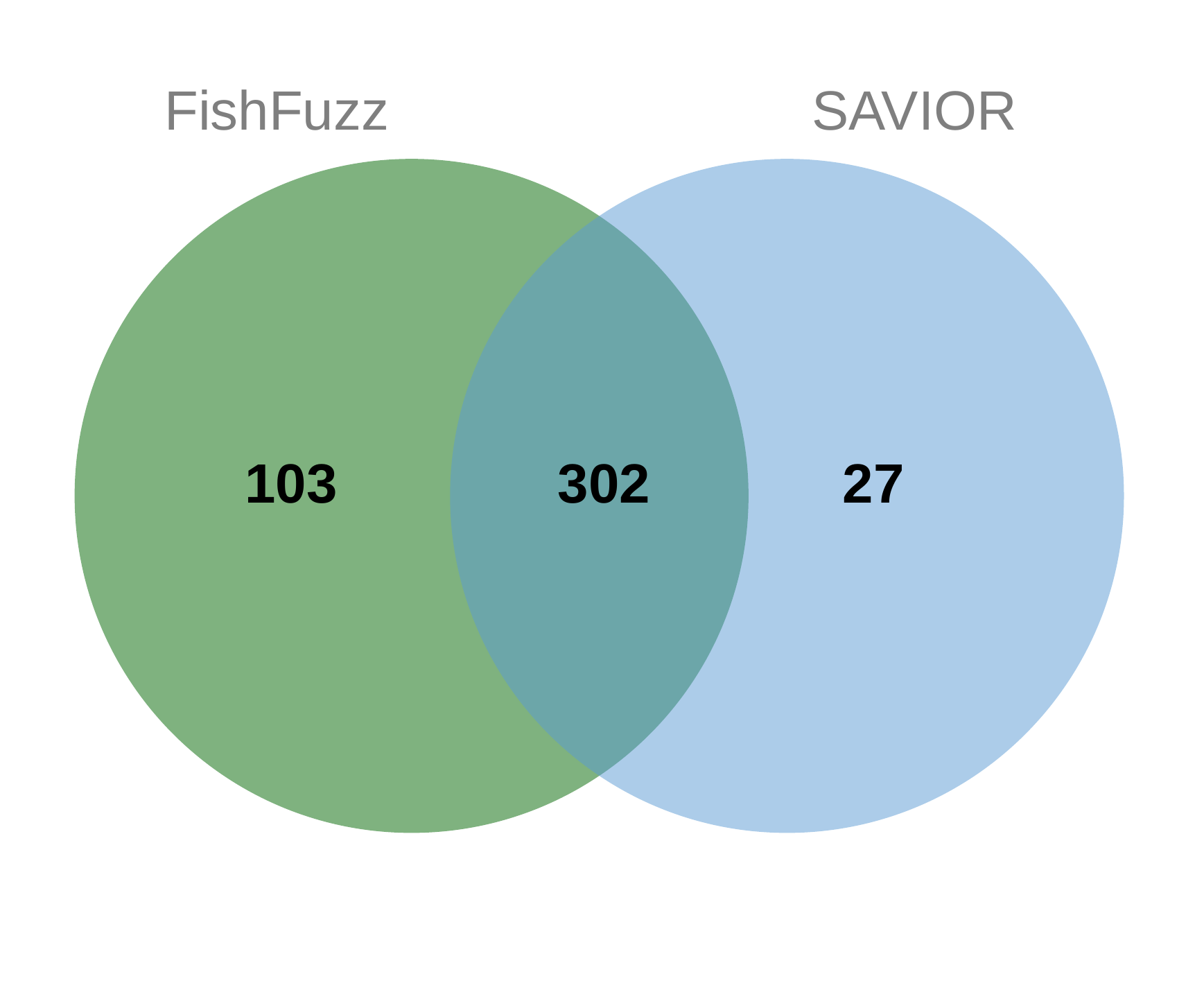}
	\caption{Triggered targets in the UBSan benchmark and compared 
		with \sys and SAVIOR. Diagram realized through~\cite{bardou2014jvenn}.}
	\label{fig:venn_ubsan}
\end{figure}

In this experiment, we want to evaluate the ability of the \emph{exploitation}
phase in triggering bugs.
To this end, we choose both ASan and UBSan benchmarks.
We note \sys is the only DGF that has been successfully deployed and tested
against two out-of-the-shell sanitizers, while previous works considered either
ASan or UBSan~\cite{osterlund2020parmesan,wang2020not,chen2020savior}.
Similar for \autoref{ssec:exploration-study}, we exercise $5$ rounds of $60$
hours each in case of ASan, while $24$ hours per run in case of UBSan (as in
the original paper~\cite{chen2020savior,osterlund2020parmesan}).
Then, we measure the number of unique bugs for ASan and the number
of triggered targets in UBSan. For ASan, we report the number of unique bugs
because each bug can be associated to multiple targets, thus it could be
ambiguous simply referring to the targets. Conversely, UBSan targets might not 
be associate to a bug, thus we prefer to indicate the targets themselves.
For instance, an integer overflow in \textsc{jasper} was considered as an
intended behavior by the authors, and thus not considered as a bug.

For ASan, we identify unique bugs by first hashing stack traces to disambiguate
crashes, followed by manually triaging bugs. For UBSan, we extract the
output pattern and identify its source location (as done by SAVIOR's authors 
after contacting them).

\paragraph*{Bugs Found}
The results for ASan and UBSan benchmarks are shown in
\autoref{tab:ASanUniq} and \autoref{tab:UBSanUniq}, respectively.
For both ASan and UBSan, \sys triggers more bugs/targets respect to the
state-of-the-art.
Specifically, \sys finds \totbugasanbest unique bugs in the ASan benchmark
for its best round, which doubles
the best rounds of AFL++ (\totbugasanbestaflp), TortoiseFuzz
(\totbugasanbesttort), and ParmeSan (\totbugasanbestparm).
We observe similar results also in terms of average unique bugs, where
\sys finds \totbugasanavg bugs against AFL++ (\totbugasanavgaflp), TortoiseFuzz
(\totbugasanavgtort), and ParmeSan (\totbugasanavgparm).
Interestingly, even though \sys reaches a lower coverage than AFL++ for
\textsc{exiv2} (\worstasancovaflp{} -- \autoref{ssec:exploration-study}), the
\emph{exploitation} phase can focus the energy and reveal more bugs.
For what concern the UBSan benchmark, we observe \sys triggers more targets
compared with SAVIOR. In particular, we activate from \minbestubsantrg to
\maxbestubsantrg more targets than SAVIOR in the best round, while from
\minavgubsantrg to \maxavgubsantrg more on average.

\paragraph*{Time-To-Exposure}
We further measure the Time-to-Exposure in \autoref{tab:ASanReal}.
Similar for the bug report, we only consider the time to exposure for the ASan benchmark
since we are more interested in bugs uniquely identifiable with an ID.
Our results show that, out of \totalbugsrepr known bugs, \sys is able to find
\totalbugsreprwin of them faster than the previous work (\totalbugsreprwinper),
\totalbugsreprdraw in the same time (\totalbugsreprdrawper),
and only \totalbugsreprloose took slightly longer time (\totalbugsreprlooseper).
Additionally, \sys find \numnewcvesfound new bugs not previously discovered, we
discuss them in a dedicated section in \autoref{ssec:finding-new-bugs}.

\paragraph*{Bugs Analysis}
Finally, we analyze the overlap of unique bugs and triggered 
targets found by \sys and its competitors.
\autoref{fig:venn_asan} and \autoref{fig:venn_ubsan} show
the intersections observed for ASan and UBSan benchmarks, respectively.
For ASan, \sys shares \numtotalbugsff unique bugs with ParmeSan, TortoiseFuzz, 
and AFL++.
We did not manage to find only \numbugsonlyaflpp bug for AFL++ and 
\numbugsonlyparm for ParmeSan, both belonging to \textsc{gpac}. We further 
investigate these cases and notice that they were triggered only once (out of 
$5$ rounds) by the respective fuzzers. We thus consider them as not
statistically relevant.
In case of UBSan, instead, we triggered \numtotaltargetsff targets in total, 
while \numtargetsonlysavior targets were triggered by SAVIOR only.
After investigation, we notice \sys did not reach the missing targets. Since 
SAVIOR is based on symbolic execution, we thus conclude SAVIOR and \sys share a 
fundamentally different exploration phase that reaches different code sections.
This leaves room for future alternative \emph{exploration} phases in \sys.

\textbf{Takeaway:} Our experiments show the ability of our
\emph{exploitation} phase to discover more bugs and trigger more targets compared
to the state-of-the-art. Additionally, we show that \sys finds known bugs
faster.

\subsection{RQ4: Can \sys find new bugs?}
\label{ssec:finding-new-bugs}

\begin{table*}[t]
    \centering
    \caption{Results of \emph{one} week of fuzzing over \numrealworldprograms
        Real-Wolrd application. We discover \numnewcvesfound confirmed CVEs,
        while
        two
        are \emph{pending}, and two are classified as \emph{only-bug}.}
    \label{tab:NewCVEs}
    \begin{tabular}{l|l|l|l|l|l|l|l}
        \toprule
        Project   & Program/Driver & Type & Sanitizer & Ref & CVE status & time
        & benchmark \\ \midrule
        libcaca   & img2txt & divide by zero & ASan & issue\_65 &
        CVE-2022-0856 & \textless{}1d & GREYONE \\
        liblouis  & lou\_checktable & heap-overflow & ASan & issue\_1171 &
        CVE-2022-26981 & \textless{}3d & GREYONE \\
		liblouis  & lou\_trace & out-of-bound read & ASan & issue\_1214 &
		CVE-2022-31783 & \textless{}3d & GREYONE \\
        tcpreplay & tcprewrite & heap-overflow & ASan & issue\_718 &
        CVE-2022-27940 & \textless{}1d & TortoiseFuzz \\
        tcpreplay & tcprewrite & assert & ASan & issue\_717& CVE-2022-27939 &
        \textless{}1d & TortoiseFuzz \\
        tcpreplay & tcpprep & heap-overflow & ASan & issue\_716 &
        CVE-2022-27941 & \textless{}1d & TortoiseFuzz \\
        tcpreplay & tcpprep & heap-overflow & ASan & issue\_719 &
        CVE-2022-27942 & \textless{}3d & TortoiseFuzz \\
        gpac & MP4Box & heap-overflow & UBSan & issue\_2138 & CVE-2022-26967
        & \textless{}3d & TortoiseFuzz \\
		gpac & MP4Box & heap-overflow & ASan & issue\_2173 & CVE-2022-29537 &
		\textless{}3d & TortoiseFuzz \\
		gpac & MP4Box & heap-overflow & ASan & issue\_2179 & CVE-2022-30976
		& \textless{}3d & TortoiseFuzz \\
		libmpeg2 & mpeg2\_dec\_fuzzer & memcpy overlap & ASan & 231026247 &
		pending & \textless{}3d & FuzzGen \\
		mujs & mujs-pp & null pointer dereference & ASan & issue\_161-1
		& bug-only & \textless{}1d & EMS \\
		mujs & mujs-pp & null pointer dereference & ASan & issue\_161-2
		& CVE-2022-30975 & \textless{}1d & EMS \\
		mujs & mujs & stack exhausted & ASan & issue\_162 & CVE-2022-30974 &
		\textless{}3d & EMS \\
		sox & sox & reachable assertion & ASan & issue\_360-1 & CVE-2022-31651 &
		\textless{}1d & MoonLight \\
		sox & sox & float pointer exception & ASan & issue\_360-2 &
		CVE-2022-31650 & \textless{}1d & MoonLight \\
        libavc & avc\_enc\_fuzzer & assert & UBSan & 223984040 & pending &
        \textless{}1d & FuzzGen \\
        ibavc & avc\_enc\_fuzzer & heap-overflow & UBSan & 224160472 & pending &
        \textless{}3d & FuzzGen \\
        Bento4    & mp4tag & heap-overflow & ASan & issue\_677 & only-bug &
        \textless{}1d & others \\
        Bento4    & mp42hevc & heap-overflow & ASan & issue\_678 &
        CVE-2022-27607 & \textless{}3d & others \\
        libsixel  & img2sixel & assert & ASan & issue\_163 & CVE-2022-27938 &
        \textless{}1d & GREYONE \\
        binutils  & nm-new & stack exhausted & ASan & 28995 & CVE-2022-27943
        & \textless{}7d & SAVIOR \\
        jasper & jasper & shift exponent exceed & ASan & issue\_311 & only-bug
        & \textless{}7d & SAVIOR \\ 
		ncurse & tic &  out-of-bound read & ASan & mail list & CVE-2022-29458 &
		\textless{}1d & GREYONE \\
		ncurse & tic & heap-overflow & ASan & mail list & bug-only &
		\textless{}1d & GREYONE \\

		\bottomrule
    \end{tabular}
\end{table*}

We challenge the ability of \sys to find new CVEs in real applications.
For this experiment, we choose \numrealworldprograms programs from top tiers
publications, \ie TortoiseFuzz~\cite{wang2020not},
SAVIOR~\cite{chen2020savior}, GREYONE~\cite{gan2020greyone},
FuzzGen~\cite{ispoglou2020fuzzgen}, as well as from the fuzzing community.
For each program, we deployed ASan and UBSan, respectively.
We run a session one week long for each program.

\autoref{tab:NewCVEs} shows the result of our experiment. In total, we found
\numnewbugsfound new bugs, \numnewcvesfound of which were confirmed CVEs.
\sys finds most of the bugs/CVEs in less than three days
(\numbuglessthanthreeday), while only \numbugmorethanthreeday required almost
seven days.
We found \numberofasanbugs bugs with the ASan sanitizers and
\numberofubsanbugs bugs with UBSan.
Specifically, most of the bugs were heap-overflow (\numdheapoverbugs). We also
found some assert violation (\numassertbugs), divide-by-zero
(\numdividezerobugs), stack-exhausted (\numstackexbugs), and shift exponential
(\numshiftexpbugs).
Interestingly, the CVE-2022-27941 was found in less than a day while previous
works tested the same program for the equivalent of $8$
weeks~\cite{wang2020not}.\footnote{The paper claims $10$ rounds of $140$ hours
each, which si around $8$ weeks.}

\textbf{Takeaway:} \sys shows to be effective in finding new CVEs since it
manages to find \numnewcvesfound new ones in less than week over programs
already deeply tested by previous works.

\subsection{RQ5: How does \sys redistribute exploration and exploitation?}
\label{ssec:phases-distribution}

In this experiment, we investigate the respective contributions of the
\emph{exploration} (both
inter- and intra-function) and \emph{exploitation} phases.
To this end, we run \sys against the UBSan benchmark for $24$ hours. Then we 
measure coverage, triggered targets, and trace the
time evolution of the \sys phases (\ie \emph{exploration} or
\emph{exploitation}).

In \autoref{fig:stages_fisfuzz}, we show covered edges correlated with
fuzzer phases. Specifically, we assign a blue background to the
\emph{inter-function exploration} (\interexplorationsquare{}), a green
background to the \emph{intra-function exploration}
(\intraexplorationsquare{}), and a red background to the \emph{exploitation
phase} (\exploitationsquare{}).
We observe two patterns.
The first pattern regards coverage-growth and \sys phases. Specifically, the
coverage tends to (statistically) grow when \sys is in one of the
\emph{exploration} phases (inter- and intra-function), while the coverage
stays steady during \emph{exploitation}.
This pattern reflects the goal of our methodology: the \emph{exploration} tends
to reach more targets (covering more code). Conversely, \sys switches to
\emph{exploitation} when it cannot reach new targets.
We can further infer this conclusion from the second pattern, where the two
\emph{exploration} phases occur more often at the beginning of the fuzzing
campaign, while the \emph{exploitation} is favored towards the end. This
again represents the design of our \emph{culling algorithm}
(\autoref{ssec:cull-queue}): when the fuzzer reaches a
coverage-wall~\cite{ispoglou2020fuzzgen} (\eg a plateau), \sys prefers the
\emph{exploitation} phase to find more bugs in already reached targets.

\newcommand{\subimagewidthxz}{0.23\textwidth}
\begin{figure}[t]
    \centering
    \begin{tabular}{p{\subimagewidthxz} p{\subimagewidthxz}}

        \includegraphics[width=\subimagewidth]{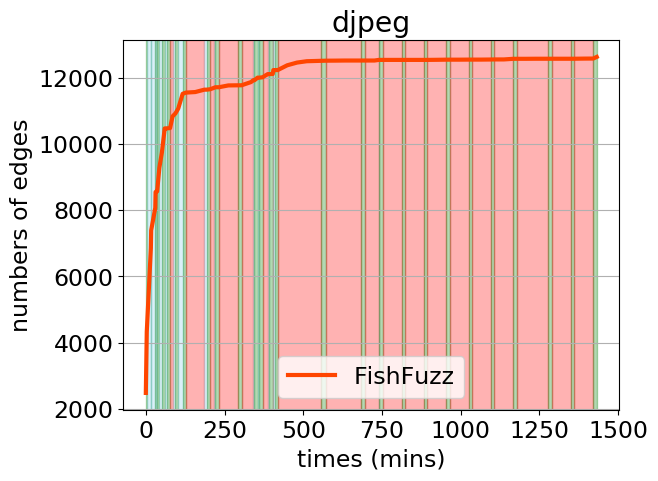} &
        \includegraphics[width=\subimagewidth]{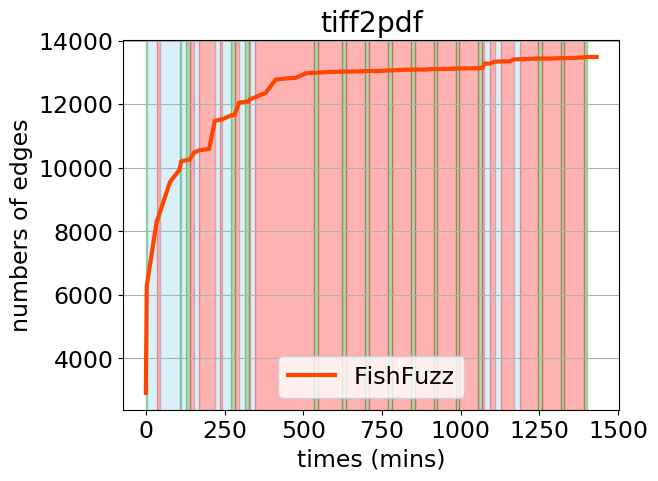} \\
        \includegraphics[width=\subimagewidth]{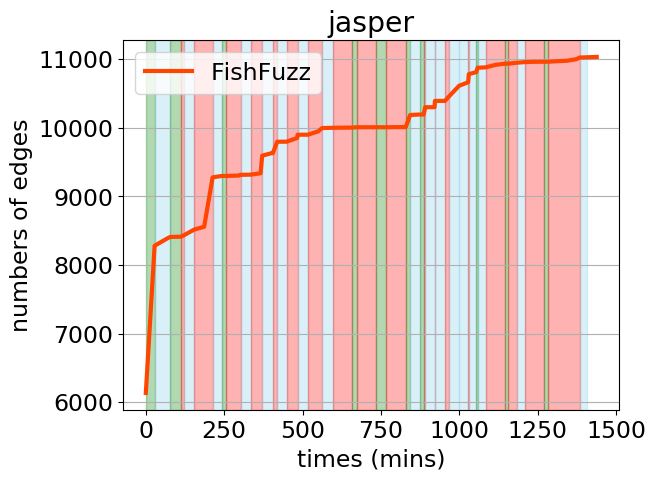} &
        \includegraphics[width=\subimagewidth]{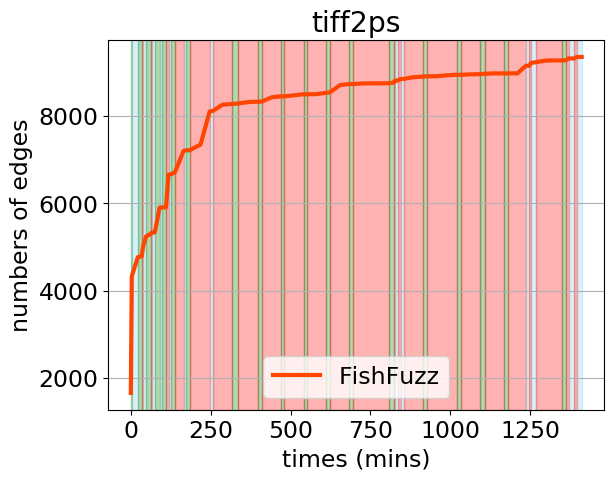} \\
        \includegraphics[width=\subimagewidth]{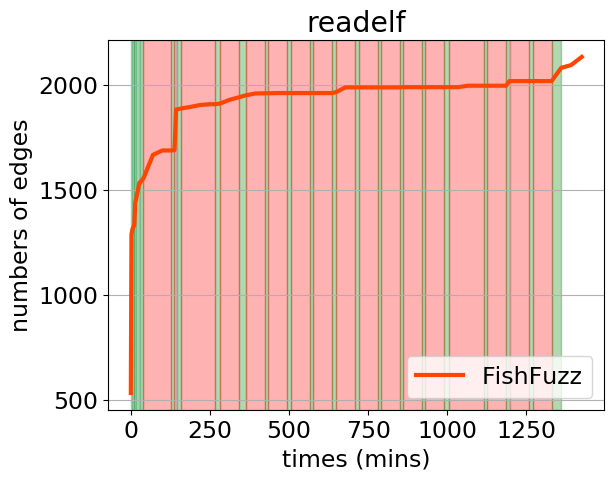} &
        \includegraphics[width=\subimagewidth]{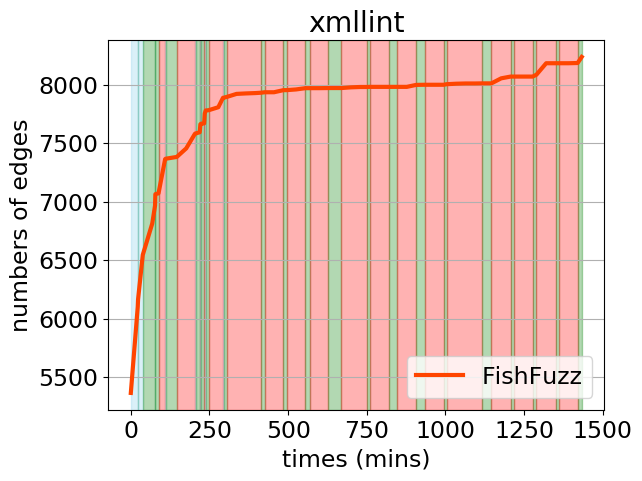} \\
        \multicolumn{2}{c}{\includegraphics[width=\subimagewidth]{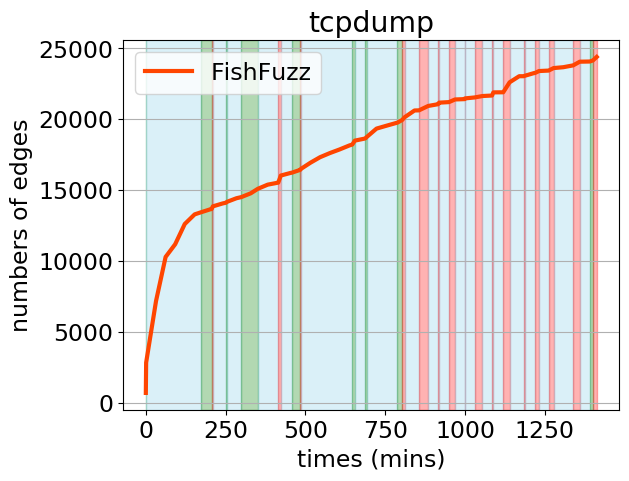}}
        \\

    \end{tabular}
    \caption{\sys stages in $24$ hour of fuzzing. \exploitationsquare{} for
    exploitation, \interexplorationsquare{} for inter-function
    exploration, and \intraexplorationsquare{} for intra-function exploration.}
    \label{fig:stages_fisfuzz}
\end{figure}

\subsection{RQ6: Can other fuzzers benefit from our strategies?}
\label{ssec:qsym-stuffs}

To answer the question if other fuzzer can profit from our strategies,
we combine QSYM~\cite{qsym} with \sys and AFL to 
measure if our cull queuing improves the performance.
Specifically, we run the QSYM with one AFL-primary and one concolic executor 
(more details in the original paper~\cite{qsym}).
Moreover, we run the experiments against the UBSan benchmark for $3$ rounds of 
$24$ hours each.

Combining QSYM+\sys improves every aspect of the original fuzzer
(\autoref{tab:QsymFish}). For instance, we improve the trigger targets up to
\qsymtrigger respect to QSYM+AFL and up to \qsymreach and \qsymcov for targets
reached and coverage, respectively.  Finally, \sys also improves the number of
seeds in the queue by reaching \qsympath more seeds at maximum (\ie path
column).

\textbf{Takeaway:} This experiment demonstrates that \sys is compositional and
helps other fuzzers improve their performance.

\begin{table*}[t]
	\centering
	\caption{Running QSYM+\sys against QSYM+AFL in the UBSan benchmark. The 
	results refer to the average of $3$ rounds for $24$ hours each. Column 
	path represents the number of seeds in the queue.}
	\label{tab:QsymFish}
	\begin{tabular}{l|rrrr|rrrr|rrrr}
	\toprule
	\multirow{2}{*}{Program} & \multicolumn{4}{c|}{QSYM+AFL} & 
	\multicolumn{4}{c|}{QSYM+\sys} & \multicolumn{4}{c}{vs 
	QSYM+AFL} \\
	  & \multicolumn{1}{l}{path} & 
	  \multicolumn{1}{l}{cov} & 
	  \multicolumn{1}{l}{reach} & \multicolumn{1}{l|}{triggered} & 
	  \multicolumn{1}{l}{path} & \multicolumn{1}{l}{cov} & 
	  \multicolumn{1}{l}{reach} & \multicolumn{1}{l|}{triggered} & 
	  \multicolumn{1}{l}{path} & \multicolumn{1}{l}{cov} & 
	  \multicolumn{1}{l}{reach} & \multicolumn{1}{l}{triggered} \\ \midrule
	djpeg    & 1304   & 10303  & 3310  & 85.7 & 2497   & 12209  & 3954  & 141.3 & \textbf{+91.54\%} & \textbf{+18.50\%} & \textbf{+19.47\%}  & \textbf{+64.98\%} \\
	jasper   & 1090   & 9730   & 1451  & 29.3 & 1778   & 11005  & 1788  & 43.3  & \textbf{+63.20\%} & \textbf{+13.10\%} & \textbf{+23.23\%}  & \textbf{+47.73\%} \\
	readelf  & 542    & 2580   & 232   & 22.0 & 666    & 2541   & 219   & 26.3  & \textbf{+22.89\%} & -1.50\%           & -5.60\%            & \textbf{+19.70\%} \\
	objdump  & 1811   & 9941   & 1068  & 62.7 & 2233   & 10048  & 1109  & 89.3  & \textbf{+23.26\%} & \textbf{+1.08\%}  & \textbf{+3.84\%}   & \textbf{+42.55\%} \\
	tcpdump  & 1741   & 12759  & 1950  & 71.7 & 2652   & 18022  & 2861  & 95.7  & \textbf{+52.37\%} & \textbf{+41.26\%} & \textbf{+46.68\%}  & \textbf{+33.49\%} \\
	tiff2pdf & 1188   & 8932   & 960   & 11.0 & 1839   & 10228  & 1277  & 12.0  & \textbf{+54.75\%} & \textbf{+14.52\%} & \textbf{+33.03\%}  & \textbf{+9.09\%}  \\
	tiff2ps  & 813    & 6181   & 496   & 7.3  & 1504   & 8365   & 1071  & 9.3   & \textbf{+84.88\%} & \textbf{+35.34\%} & \textbf{+116.07\%} & \textbf{+27.27\%} \\
	xmllint  & 2232   & 8103   & 672   & 9.3  & 2741   & 8561   & 734   & 10.7  
	& \textbf{+22.81\%} & \textbf{+5.65\%}  & \textbf{+9.17\%}   & 
	\textbf{+14.29\%} \\ \midrule
	total    & 10720  & 68528  & 10139 & 299.0  & 15909  & 80980  & 13012 & 
	428.0 & \textbf{+48.40\%} & \textbf{+18.17\%} & \textbf{+28.34\%}  & 
	\textbf{+43.14\%} \\ \bottomrule
	\end{tabular}
\end{table*}

\section{Discussion}
\label{sec:discussion}

Here, we discuss limitations and future work for \sys.
Specifically, we focus on performance (\autoref{ssec:speed}),
target size (\autoref{ssec:target-set-size}), and combination with
orthogonal techniques (\autoref{ssec:pruning-unfeasible-paths}).

\subsection{Performance}
\label{ssec:speed}

Our prototype extends AFL with
instrumentation to trace functions and targets information.
During our extensive
experiments,
we notice our prototype does not achieve optimal performances in some cases 
(\textsc{exiv2} in \autoref{tab:ASanCov}), adversely affecting the fuzzing 
campaign. We tracked the root cause to two problems:
First, AFL++ uses more efficient instrumentation compared to AFL.
Second, \sys uses additional shared memory to trace explored functions
and reached targets. Synchronizing with this structure introduces latency and 
reduces the number of seeds exercised.
To overcome these limitations, we are currently porting our prototype to
AFL++ to leverage the more efficient instrumentation.
We realized this limitation during our evaluation against AFL++
(\autoref{tab:ASanCov}).
With this (engineering) optimization, \sys performance will further improve and
we will integrate those results.

\subsection{Target Set Size}
\label{ssec:target-set-size}

\sys is designed to handle large target sets (up to \maxtargetscovered
in our experiments \autoref{sec:evaluation}).
Even though \sys scales up efficiently in these cases, we also observe a drop
of performances for small target sets (\eg at around tens targets).
We plan to tackle this problem in two directions.
First, we could employ different mutators to direct seeds faster, such
as~\cite{lyu2019mopt}.
Second, we believe this observation suggests the need of specific ad-hoc 
seed-distance metrics according to the context.
Therefore, we will investigate the performance of different
seed-target metrics and infer the best trade-off as future work.

\subsection{Combining \sys with other works}
\label{ssec:pruning-unfeasible-paths}

BEACON~\cite{huangbeacon} (to be published at Oakland'22), a concurrent DGF uses
software analysis (\eg static or dynamic) to foresee (and discard) unreachable
portions of code. \sys would benefit from these techniques to speed up the
initial exploration phase or better re-assign energy to targets.  Similarly, we
consider to combine SAVIOR~\cite{chen2020savior} with our \emph{exploration}
phase to investigate if different approaches can lead to better performances.

\section{Related Works}

\sys improves existing fuzzing work across two research areas:
Directed Greybox Fuzzers (\autoref{ssec:dgf-rw}) and Multistage
Fuzzers (\autoref{ssec:explexpl-rw}).

\subsection{Directed Greybox Fuzzers}
\label{ssec:dgf-rw}

DGF is a branch of fuzzing that specializes
fuzzers for hitting a given set of targets (instead of improving code-coverage).

B{\"o}hme et al. discusses the first prototype, AFLGo~\cite{bohme2017directed},
which models the distance seed-targets as an harmonic average distance.
However, the AFLGo approach losses precision for large target sets.
In this regard, \sys relies on a novel seed-target distance whose precision is
not affected by the number of targets.
Improvements to AFLGo were further proposed by Chen et al. with
Hawkeyes~\cite{chen2018hawkeye} and Peiyuan et al. with
FuzzGuard~\cite{fuzzguard}.
These works try to handle indirect calls by adopting heavy weight static
analysis (Hawkeyes) or using deep learning to discard unfruitful inputs
(FuzzGuard), respectively.
Conversely, \sys does not need any complex analysis to resolve indirect jumps,
while its seed selection automatically promotes interesting inputs.

Steps toward more scalable DGF are discussed by {\"O}sterlund with
ParmeSan~\cite{osterlund2020parmesan} and Chen with
SAVIOR~\cite{chen2020savior}, respectively.
Both ParmeSan and SAVIOR consider as targets all the sanitizers labels.
Additionally, SAVIOR introduces an heavy reachable analysis to select
interesting inputs.
Both works suffer from the original AFLGo limitation since they collapse the
distance seed-targets into a scalar, thus loosing precision.
\sys differs from these works for two reasons: first, it employs a novel
distance seed-targets that overcomes scalability limitations, second, it uses
a faster exploration phase to boost the targets discovery.

Gwangmu et al. propose CAFL~\cite{cafl} (Constraint guided directed greybox
fuzzing). The goal of this work is to synthesis a POC from a given crash by
following a similar approach of AFLGo.
In their scenario, CAFL considers only one target, while \sys is designed to
handle a large number of targets.
Finally, Xiaogang et al. discuss Regression Greybox Fuzzing~\cite{regression}, 
their work contains methods to select possible bogus code locations by
analyzing the repository history.
This approach is then combined with a more efficient power schedule policy.
We consider this work as orthogonal to \sys since we focus on the seed
selection strategy, while they recognize interesting code locations for testing.

Huange et al. introduces BEACON~\cite{huangbeacon}, which uses sophisticated
static-analysis to remove unfeasible paths, thus speeding up the
\emph{exploration} phase. Conversely, \sys aims at improving the exploitation
phase and trigger targets. We consider their approach as orthogonal to \sys, we
further plan to combine the two strategies in the future.

\subsection{Multistage Fuzzers}
\label{ssec:explexpl-rw}

Multistage fuzzers use exploration and exploitation phases to reach and trigger
multiple targets.
B{\"o}hme et al. proposes AFLFast~\cite{aflfast}, which relies on Markov chain
to probabilistically select seeds that improve the coverage. Their contribution
is mainly energy distribution related, while queue culling and seed distance
are not discussed.

Lemieux et al.~\cite{fairfuzz} study new mutation strategies and seed
selections to hit rare branches.
Their contribution is more related to improve code-coverage, while \sys also
maximizes the targets triggered.
Yue et al. discuss a combination of adaptive energy schedule and game theory to
avoid testing unfruitful seeds. Their approach does not discuss seeds
selection strategies, thus being orthogonal to \sys.

Wang et al.~\cite{wang2020not} select interesting targets upon extensive
software analysis, that are then combined with a novel queue culling strategy.
However, their approach considers only time-invariant targets, thus not
adapting the fuzzer energy toward more promising code locations.
Conversely, the queue culling mechanism of \sys is adaptive and can be
potentially used to improve the performances of Wang's work.

\section{Conclusion}

Directed Greybox Fuzzing has been hampered by averaged distance metrics that
over-eagerly aggregate paths into scalars and simple energy distribution that
simply assigns equal energy to all targets in a round robin fashion.

We draw inspiration from trawl fishing where a wide net is cast and pulled to
reach many targets before they are harvested.
\sys improves the \emph{exploration} and \emph{exploitation} phases with
explicit feedback for both phases and a dynamic switching strategy that
alternates mutation and energy distribution based on the current phase.
Additionally, our dynamic target ranking automatically discards exhausted 
targets and our novel multi-distance metric keeps track of tens of thousands of 
targets without loss of precision.

We evaluate \sys against \numtotalprograms programs and have, so far, discovered
\numnewbugsfound new bugs (\numnewcvesfound CVEs). \sys will be released as open
source and we provide a test environment to play with our novel DGF.


\bibliographystyle{IEEEtran}
\bibliography{FishFuzz}

\end{document}